\documentclass[numberedappendix,12pt,preprint]{emulateapj}

\usepackage[]{amsmath}

\shorttitle{EoR Inference from \lya}
\shortauthors{Mason et al. (2018)}

\usepackage{color}			   
\definecolor{midgray}{gray}{0.4}		
\definecolor{orange}{rgb}{1,0.5,0}  
\usepackage[shortlabels]{enumitem}

\usepackage[colorlinks=true,citecolor=blue,linkcolor=blue]{hyperref}    

\usepackage{etoolbox}
\makeatletter
\patchcmd{\NAT@citex}
  {\@citea\NAT@hyper@{\NAT@nmfmt{\NAT@nm}\NAT@date}}
  {\@citea\NAT@nmfmt{\NAT@nm}\NAT@hyper@{\NAT@date}}
  {}
  {}
\patchcmd{\NAT@citex}
  {\@citea\NAT@hyper@{%
     \NAT@nmfmt{\NAT@nm}%
     \hyper@natlinkbreak{\NAT@aysep\NAT@spacechar}{\@citeb\@extra@b@citeb}%
     \NAT@date}}
  {\@citea\NAT@nmfmt{\NAT@nm}%
   \NAT@aysep\NAT@spacechar%
   \NAT@hyper@{\NAT@date}}
  {}
  {}
\patchcmd{\NAT@citex}
  {\@citea\NAT@hyper@{%
     \NAT@nmfmt{\NAT@nm}%
     \hyper@natlinkbreak{\NAT@spacechar\NAT@@open\if*#1*\else#1\NAT@spacechar\fi}%
       {\@citeb\@extra@b@citeb}%
     \NAT@date}}
  {\@citea\NAT@nmfmt{\NAT@nm}%
   \NAT@spacechar\NAT@@open\if*#1*\else#1\NAT@spacechar\fi%
   \NAT@hyper@{\NAT@date}}
  {}
  {}
\makeatother

\newcommand{\simgt}{\,\rlap{\lower 3.5 pt \hbox{$\mathchar \sim$}} \raise
1pt \hbox {$>$}\,}
\newcommand{\simlt}{\,\rlap{\lower 3.5 pt \hbox{$\mathchar \sim$}} \raise
1pt \hbox {$<$}\,}

\newcommand{\Ob}{\Omega_{\textrm b}}
\newcommand{\Om}{\Omega_{\textrm m}}
\newcommand{\OL}{\Omega_{\Lambda}}

\newcommand{\msun}{M_{\odot}}

\newcommand{\lya}{Ly$\alpha$}

\newcommand{\xHI}{{\overline{x}_\textrm{\scriptsize{\textsc{hi}}}}}
\newcommand{\DV}{{\Delta v}}
\newcommand{\Tigm}{{\mathcal{T}_\textsc{igm}}}
\newcommand{\MUV}{M_\textsc{uv}}
\newcommand{\HI}{H \textsc{i}}
\newcommand{\HII}{H \textsc{ii}}

\DeclareMathOperator\erfc{erfc}

\newcommand{\BE}{\begin{equation}}
\newcommand{\EE}{\end{equation}}
\newcommand{\BEA}{\begin{eqnarray}}
\newcommand{\EEA}{\end{eqnarray}}

\begin{document}

\title{The Universe is Reionizing at $z\sim7$: Bayesian Inference of \\ the IGM Neutral Fraction Using \lya\ Emission from Galaxies}

\author{
Charlotte A. Mason$^{1}$,
Tommaso Treu$^{1}$,
Mark Dijkstra$^{2}$,
Andrei Mesinger$^{3}$,
Michele Trenti$^{4,5}$,\\
Laura Pentericci$^{6}$,
Stephane de Barros$^{7,8}$, and
Eros Vanzella$^{8}$,
}
\affil{$^{1}$ Department of Physics and Astronomy, UCLA, Los Angeles, CA, 90095-1547, USA}
\affil{$^{2}$ Institute of Theoretical Astrophysics, University of Oslo, P.O. Box 1029, N-0315 Oslo, Norway}
\affil{$^{3}$ Scuola Normale Superiore, Piazza dei Cavalieri 7, I-56126 Pisa, Italy}
\affil{$^{4}$ School of Physics, University of Melbourne, Parkville, Victoria, Australia}
\affil{$^{5}$ ARC Centre of Excellence for All Sky Astrophysics in 3 Dimensions (ASTRO 3D)}
\affil{$^{6}$ INAF Osservatorio Astronomico di Roma, Via Frascati 33, I-00040 Monteporzio (RM), Italy}
\affil{$^{7}$ Observatoire de Gen\`{e}ve, Universit\`{e} de Gen\`{e}ve, 51 Ch. des Maillettes, 1290 Versoix, Switzerland}
\affil{$^{8}$ INAF Osservatorio Astronomico di Bologna, via Ranzani 1, 40127 Bologna, Italy}
\email{cmason@astro.ucla.edu}

\begin{abstract}
We present a new flexible Bayesian framework for directly inferring the fraction of neutral hydrogen in the intergalactic medium (IGM) during the Epoch of Reionization (EoR, $z \sim 6 - 10$) from detections and non-detections of Lyman Alpha (\lya) emission from Lyman Break galaxies (LBGs). Our framework combines sophisticated reionization simulations with empirical models of the interstellar medium (ISM) radiative transfer effects on \lya. We assert that the \lya\ line profile emerging from the ISM has an important impact on the resulting transmission of photons through the IGM, and that these line profiles depend on galaxy properties. We model this effect by considering the peak velocity offset of \lya\ lines from host galaxies' systemic redshifts, which are empirically correlated with UV luminosity and redshift (or halo mass at fixed redshift). We use our framework on the sample of LBGs presented in \citet{Pentericci2014} and infer a global neutral fraction at $z\sim7$ of $\xHI = 0.59_{-0.15}^{+0.11}$, consistent with other robust probes of the EoR and confirming reionization is on-going $\sim700$ Myr after the Big Bang. We show that using the full distribution of \lya\ equivalent width detections and upper limits from LBGs places tighter constraints on the evolving IGM than the standard \lya\ emitter fraction, and that larger samples are within reach of deep spectroscopic surveys of gravitationally lensed fields and JWST NIRSpec.
\end{abstract}

 
\section{Introduction}
\label{sec:intro}

In the first billion years of the universe's history, intergalactic hydrogen atoms, formed at Recombination, were ionized \citep[e.g.,][]{Robertson2015,Mesinger2016a,PlanckCollaboration2016a}. This reionization of the intergalactic medium (IGM) was driven by the first sources of light: stars, and accretion disks around black holes, in galaxies. By understanding the process and timeline of reionization we can learn about the nature of these nascent populations of galaxies. 

Ground-breaking observations within the last decade have provided significant information about this {\em Epoch of Reionization} (EoR, $z\sim6-10$). With the largest near-IR instruments in space and on the ground we have now discovered large populations of galaxies at $z\simgt6$ \citep[e.g.,][]{McLure2009,Trenti2011,Bradley2012,Illingworth2013,Schenker2013,Schmidt2014a,Yue2014,Bouwens2015a,Finkelstein2015,Calvi2016}. Young stars in these galaxies are likely the primary sources of reionizing photons \citep[e.g.,][]{Lehnert2003,Bouwens2003,Yan2003,Bunker2004,Finkelstein2012,Robertson2013,Schmidt2014a}, though a contribution from AGN cannot be excluded \citep{Giallongo2015,Madau2015,Onoue2017}: we do not know if sufficient hard ionizing photons escape from galaxies as we do not fully understand the interactions between these early galaxies and their surrounding media.

Absorption features in quasar spectra suggest reionization was largely complete by $z\sim6$ \citep[$<1$ Gyr after the Big Bang, e.g.,][]{Fan2006,Schroeder2013,McGreer2014,Venemans2015}, whilst the electron scattering optical depth to the CMB \citep{PlanckCollaboration2015,PlanckCollaboration2016a,Greig2016} indicates significant reionization was occurring at $z\sim7.8-8.8$. A robust constraint, albeit from a single sightline, on on-going reionization comes from the absorption spectrum of the $z=7.1$ quasar ULAS J1120+0641, where \citet{Greig2016b} recently inferred a neutral fraction of $\xHI = 0.40_{-0.19}^{+0.21}$.

To produce a timeline of reionization consistent with the evolution suggested by observations generally requires optimistic assumptions about the numbers of as-yet undetected ultra-faint galaxies \citep{MichaelShull2012,Robertson2013,Mason2015a} -- which are likely the hosts of high redshift gamma-ray bursts \citep{Kistler2009,Trenti2012a}, and/or the production efficiency and escape fraction of hard ionizing photons \citep{Bouwens2015c,Ma2015,Grazian2017,Vanzella2017}. However, the timeline of reionization is not well-constrained, especially beyond $z\simgt6$ where quasars become extremely rare \citep{Fan2001,Manti2016}.

Into the EoR, a powerful probe of the IGM is the Lyman alpha (\lya, 1216\AA) emission line from galaxies, which is strongly attenuated by neutral hydrogen \citep{Haiman1999,Malhotra2004,Santos2004,Verhamme2006,McQuinn2007,Dijkstra2014a}. Observing \lya\ at high redshift gives us key insights into both the IGM ionization state and galaxy properties, and, whilst quasars probably live in the densest regions of the early universe \citep{Mesinger2010}, observing galaxies enables us to trace reionization in cosmic volumes in a less biased way.

Dedicated spectroscopic follow-up of young star-forming galaxies at high redshift, identified as photometric dropouts (Lyman break galaxies, hereafter LBGs) combined with low redshift comparison samples \citep{Hayes2013,Yang2016} show that the fraction of LBGs emitting \lya\ increases with redshift \citep{Stark2010,Curtis-Lake2012,Hayes2011,Cassata2015}, likely because the dust fraction in galaxies decreases \citep[e.g.,][]{Finkelstein2012a,Bouwens2013} which reduces the absorption of \lya\ \citep{Hayes2011}. However, there is a potential smoking gun signature of reionization at $z>6$: recent observations show a declining fraction of \lya\ emitters in the LBG population with redshift \citep[e.g.][]{Fontana2010a,Stark2010,Caruana2012,Treu2013,Caruana2014,Faisst2014,Tilvi2014,Schenker2014,Pentericci2014}, as well as an evolving \lya\ luminosity function \citep[e.g.,][]{Ouchi2010,Zheng2017a,Ota2017,Konno2017}, suggesting an increasingly neutral, but inhomogeneous, IGM \citep{Dijkstra2014,Mesinger2014}.

Robust conversions from observations to the IGM state are challenging, however, and current constraints from \lya\ emission measurements show some tension. The sudden drop in \lya\ emission from LBGs suggests a high neutral fraction at $z\sim7$, $\xHI \simgt 0.5$ \citep{Dijkstra2014,Choudhury2015,Mesinger2014}, whereas measurements from clustering of \lya\ emitters at $z=6.6$ imply a lower neutral fraction \citep[$\xHI \simlt 0.5$,][]{Ouchi2010,Sobacchi2015a}. These constraints are consistent within 1$\sigma$ but the qualitative tension motivates a more thorough treatment of the properties of \lya\ emitters during reionization. Given this, and that tight constraints on the reionization history can constrain properties of the sources of reionization \citep[e.g., the minimum mass/luminosity of galaxies, and the escape fraction of ionizing photons,][]{Bouwens2015b,Mitra2016,Greig2017,Greig2016}, we aim to develop a robust framework for inferring the ionization state of the IGM from observations \lya\ from galaxies.

The conversion from the evolving transmission of \lya\ emission from galaxies to a constraint on the IGM ionization state is non-trivial and involves physics from pc to Gpc scales. Multiple observations \citep[e.g.,][]{Treu2013,Pentericci2014,Becker2015} and simulations \citep{Furlanetto2006a,McQuinn2007a} suggest reionization of the IGM is likely a `patchy' process, with large ionized bubbles growing faster in overdense regions filled with star-forming galaxies. An accurate model of reionization must include realistic large-scale IGM structure \citep{Trac2008,Iliev2014,Sobacchi2014}. 

Irrespective of reionization, as a highly resonant line, \lya\ photons experience significant scattering within the interstellar medium (ISM) of their host galaxies, and absorption within the circumgalactic medium (CGM) which affects the visibility of emission \citep{Verhamme2006,Verhamme2008,Dijkstra2007,Laursen2011}. ISM effects on \lya\ are likely to correlate with galaxy mass and star formation rate (SFR) via dust absorption, neutral hydrogen column density and covering fraction, and outflows \citep{Erb2014,Erb2015,Oyarzun2016,Hayward2015,Yang2017}.

UV faint galaxies ($\MUV > M^* \sim -20$) tend to be the strongest \lya\ emitters at all redshifts due to lower dust masses and neutral hydrogen column densities \citep{Yang2016,Yang2017}. However, a small sample of UV bright galaxies at $z > 7.5$ with strong Spitzer/IRAC excesses have recently been observed with \lya\ \citep{Finkelstein2013,Roberts-Borsani2015,Oesch2015,Zitrin2015a,Stark2017}, at a redshift when the IGM is expected to be significantly neutral \citep{PlanckCollaboration2016a,Greig2016}. Are these objects a new class of highly ionizing galaxies \citep{Stark2017}, emitting \lya\ with very high EW so some flux is still observable even after attenuation in the IGM? Do they inhabit large ionized bubbles in the IGM at high redshift? How do different halo environments and ISM properties affect the impact on reionization on galaxies?

\citet{Dijkstra2011} first considered the effects of the ISM on \lya\ photons during reionization, using shell models \citep[e.g.][]{Verhamme2006,Gronke2015a} to mimic the ISM radiative transfer, and showed ISM effects had a large impact on the transmission of \lya\ photons through the reionizing IGM. As described above, the \lya\ photons' journey through the ISM depends on galaxy properties. However, previous constraints on the evolving transmission of \lya\ emission at $z\simgt6$ have limited treatment of this effect: \citet{Dijkstra2011} and \citet{Mesinger2014} parametrically accounted for the ISM but assumed the LBG galaxy population is homogeneous; \citet{Jensen2012} obtained similar results combining cosmological hydro-simulations of reionization with a different sub-grid prescription for \lya\ radiative transfer in the ISM; simpler models do not treat the ISM but consider two bins of UV bright and faint galaxies \citep[e.g.,][]{Treu2012}. 

In this paper we introduce a flexible modeling framework to enable Bayesian inference of the IGM neutral fraction from detections and non-detections of \lya\ from LBGs. Our framework includes realistic cosmological IGM simulations which contain the large-scale structure of the reionizing IGM. We generate 1000s of sightlines through these simulations to halos born from the same density field as the IGM, and populate these halos with simple, but realistic, ISM properties drawn from empirical models, which, for the first time in a reionization model, are linked to observable galaxy properties.

Our model asserts the impact of the ISM on the \lya\ line profile is the most important galaxy property to consider when trying to make accurate inferences about reionization. In our model we include this effect via the peak velocity offset of the \lya\ line profile from systemic ($\DV$), which correlates with galaxy mass (or UV magnitude at fixed redshift), for which there are a handful of measurements at $z \simgt 6$ \citep{Pentericci2016,Bradac2016,Mainali2016,Stark2017}. Galaxies with high \lya\ velocity offsets have higher probabilities of transmitting \lya\ photons through the IGM. This effect is robustly accounted for in our model as a nuisance parameter in our inference.

The paper is structured as follows: in Section~\ref{sec:lyart} we explain the ISM, CGM, and IGM radiative transfer modeling components of our model; in Section~\ref{sec:bayes} we describe our flexible Bayesian framework for inferring the neutral fraction $\xHI$; in Section~\ref{sec:results} we give our results including key insights from the model, the inferred value of $\xHI$ from current observations and forecasts for spectroscopic surveys with the James Webb Space Telescope (JWST); we discuss our results in Section~\ref{sec:dis} and present a summary and conclusions in Section~\ref{sec:conc}.

We use the \citet{PlanckCollaboration2015} cosmology where $(\OL, \Om, \Ob, n,  \sigma_8, H_0) =$ (0.69, 0.31, 0.048, 0.97, 0.81, 68 km s$^{-1}$ Mpc$^{-1}$), and all magnitudes are given in the AB system.

\section{ISM, CGM, and IGM Radiative Transfer Modeling}
\label{sec:lyart}
\lya\ photons are significantly affected by the neutral hydrogen they encounter within the ISM of their source galaxies, their local CGM, and the IGM through which they travel to our telescopes. To make constraints in the Epoch of Reionization we must model \lya\ radiative transfer in all three media. Here we describe the combination of empirical formalisms and numerical simulations to model the effect of the ISM (Section~\ref{sec:lyart_ISM}) and the CGM and IGM (Section~\ref{sec:lyart_IGM}) on \lya\ transmission.

\subsection{ISM \lya\ radiative transfer}
\label{sec:lyart_ISM}

\lya\ photons are produced predominantly via recombination in \HII\ regions around young stars and have a high cross-section for resonant scattering \citep[for a detailed review see][]{Dijkstra2014a}. As the ISM of individual galaxies contains a large amount of neutral hydrogen gas to escape the ISM \lya\ photons must diffuse both spatially \textit{and} spectrally \citep[e.g.,][]{Shapley2003,McLinden2011,Chonis2013,Mostardi2013,Song2014}. This produces the fiducial double-peaked \lya\ lineshape, for which the red (blue) peak is enhanced for outflows (inflows) \citep{Zheng2002,Verhamme2006}.

\begin{figure}[t] 
\includegraphics[width=0.49\textwidth]{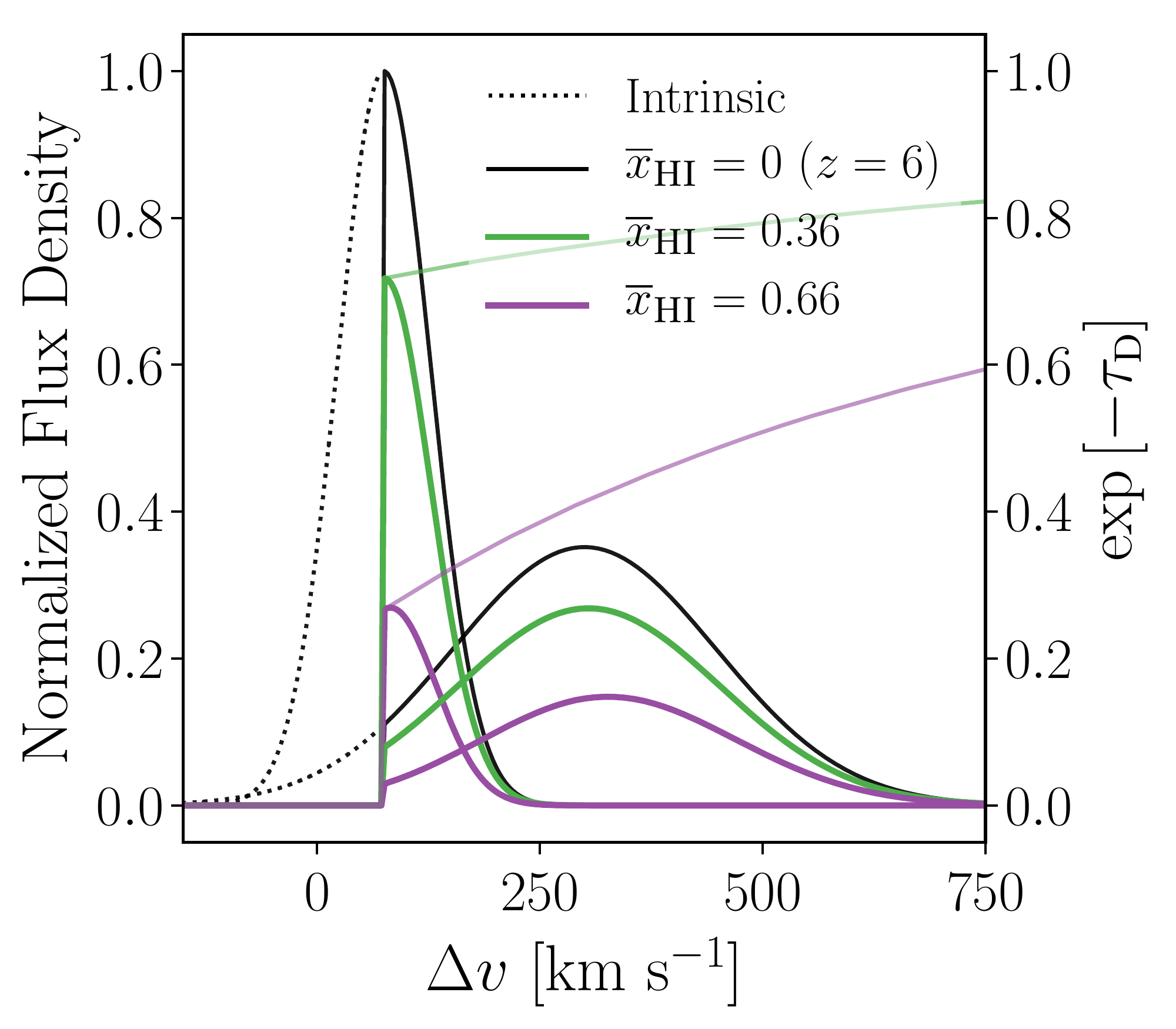}
\caption{The effect of the IGM on simulated line profiles. We show two example intrinsic line profiles (black dotted lines), with peak velocity offsets of 75 and 300 km s$^{-1}$, with flux densities normalized to that of the line at 75 km s$^{-1}$. This is the line \textit{after transmission through the ISM}. The solid black shows the lineshape in an ionized universe at $z\sim6$ where \textit{all flux bluer than the halo's circular velocity is resonantly absorbed} by neutral hydrogen in the local CGM/IGM (i.e. they experience only $\tau_\mathrm{\HII}$). The colored lines show the emission lines after transmission through a reionizing IGM with damping wing optical depths $\tau_\textsc{d}$, where the median IGM attenuation is also plotted (lighter line, corresponds to right axis). Lines emitted with high velocity offsets are less attenuated by the IGM: for the line with $\DV = 75$ km s$^{-1}$ $\sim70\%$ of the emitted flux is observed for $\xHI=0.36$ (green), for the line with $\DV = 300$ km s$^{-1}$ this fraction rises to $\sim75\%$. For $\xHI=0.66$ (purple) $\sim30\%$ of the total flux is transmitted from the line with $\DV=75$ km s$^{-1}$ whilst $\sim40\%$ is emitted for the line at $\DV=300$ km s$^{-1}$.}
\label{fig:lineprofile}
\end{figure}

In this work, we model the \lya\ lineshape \textit{after transmission through the ISM} as a Gaussian, centered at a velocity offset $\DV$ from the systemic redshift of the galaxy (due to scattering through the ISM, described in Section~\ref{sec:lyart_ISM_mhdv}) with a velocity dispersion $\sigma_\alpha$ (due to scattering and thermal broadening in the ISM, described in Section~\ref{sec:lyart_ISM_sigma}). We refer to this lineshape as `intrinsic', examples are shown as dotted black lines in Figure~\ref{fig:lineprofile}. As described below in Section~\ref{sec:lyart_IGM} even after reionization residual neutral gas in the IGM and CGM will absorb all blue flux at $z\simgt6$.

\subsubsection{Modeling \lya\ velocity offsets}
\label{sec:lyart_ISM_mhdv}

\begin{figure*}[t] 
\includegraphics[width=0.99\textwidth]{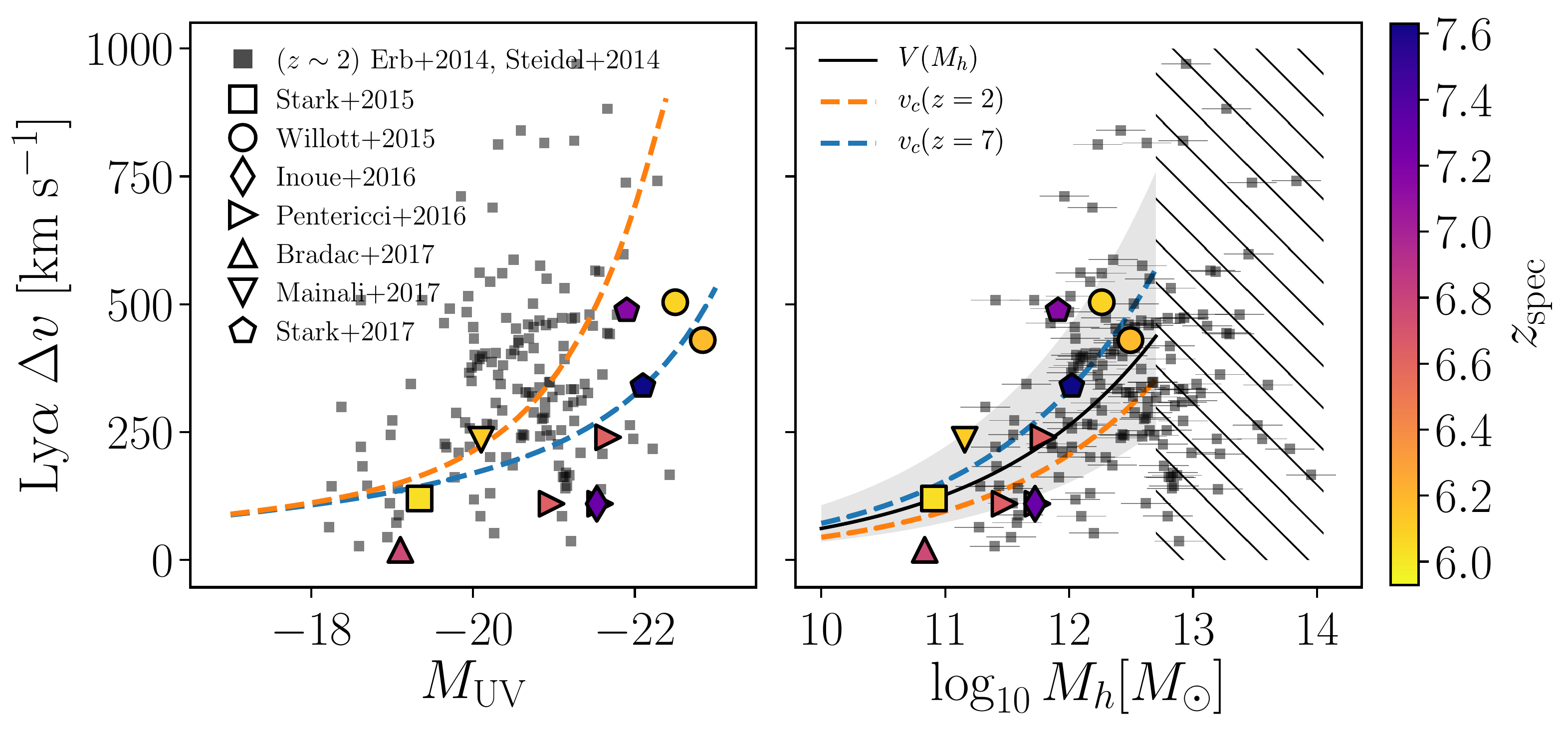}
\caption{\lya\ velocity offset as a function of UV absolute magnitude (left), halo mass (right, derived from the \citet{Mason2015a} UV magnitude - halo mass relation) for a collection of data from the literature \citep{Erb2014,Steidel2014,Willott2015,Stark2015,Stark2017,Inoue2016,Mainali2016,Pentericci2016,Bradac2016}. The gray squares show data from a $z\sim2$ sample, whilst the colored points are at $z>6$. We take the $z\sim2$ distribution as complete and intrinsic and fit a log-normal distribution to the $\DV - M_h$ points as shown in Equation~\ref{eqn:pDV_Mh}. The median $\DV - M_h$ fit is shown as a black solid line, and the gray shaded region shows the $\sigma_v$ scatter. We add a 0.2 mag uncertainty to the UV magnitude measurements to account for scatter in the UV magnitude -- halo mass relation and use the propagated uncertainties in halo mass in the $\DV - M_h$. The hashed region in the right panel indicates the galaxies with $\MUV < -21$ which are discarded from fitting due to large uncertainties in assigning their halo masses. We plot the circular velocities, $v_c$ of halos at $z\sim2$ (dashed orange) and $z\sim7$ (dashed blue) for comparison. The $\DV - M_h$ relation closely traces the circular velocities, suggesting galaxy mass is a key mediator of \lya\ radiative transfer.}
\label{fig:DV_Muv_lit}
\end{figure*}

Numerous studies of star-forming galaxies at $z\simlt4$ have identified the column density of neutral hydrogen ($N_\textsc{hi}$) within the ISM as a key mediator of \lya\ radiative transfer. \lya\ photons traveling through highly dense neutral ISM scatter more frequently and emerge with larger velocity offsets than galaxies with lower $N_\textsc{hi}$ \citep{Shibuya2014a,Hashimoto2015,Yang2016,Yang2017,Guaita2017}. 

Low mass galaxies, especially at high redshifts, are less likely to contain significant fractions of neutral gas due to enhanced photoionization feedback. Additionally, strong star formation feedback may drive outflows and/or reduce the covering fraction of neutral gas in the ISM which can facilitate \lya\ escape \citep{Jones2013,Trainor2015,Leethochawalit2016}. 

Recently, a correlation has been suggested between UV magnitude and \lya\ velocity offset \citep{Schenker2013a,Erb2014,Song2014,Stark2015,Mainali2016,Stark2017}, again indicating galaxy mass and/or SFR strongly affects \lya\ escape. However, galaxies with the same UV magnitudes at different redshifts likely have very different masses because of increasing SFR at high redshift \citep[e.g.,][]{Behroozi2013a,Barone-Nugent2014,Mason2015a,Finkelstein2015b,Harikane2015,Harikane2017} so one should be cautious of comparing galaxies with the same UV magnitudes at different redshifts. We plot a compilation of $\MUV - \DV$ measurements from the literature \citep{Erb2014,Steidel2014,Bradac2016,Inoue2016,Mainali2016,Pentericci2016,Stark2015,Stark2017,Willott2015} in Figure~\ref{fig:DV_Muv_lit} (left) where it is clear the high redshift galaxies have lower $\DV$ at given $\MUV$ compared to the low redshift galaxies, probably because they have lower mass.

To model the effect of the ISM on \lya\ escape we assume the column density of neutral hydrogen within the ISM is determined by halo mass and is the most important quantity for understanding the emerging \lya\ line profile. This is likely an over-simplification, e.g. `shell' models take $\sim6$ parameters to model \lya\ lines \citep{Verhamme2006,Verhamme2008,Gronke2016}, but is an efficient first-order approach. With this in mind, we assume a correlation between $\DV$ and halo mass of the form $\DV \sim (M_h)^m$, where we determine $m$ empirically from observations, as described below.

We take a sample of 158 $z\sim2-3$ galaxies with both UV magnitudes and \lya\ velocity offsets from \citet{Erb2014} and \citet{Steidel2014}. The \citet{Steidel2014} sample (from the KBSS-MOSFIRE survey) is effectively complete at $\MUV \simlt -20$ where $\sim90\%$ of their photometrically-selected LBGs have rest-frame optical emission lines detected in deep near-IR spectroscopy with \textit{Keck}/MOSFIRE \citep{McLean2012}. The \citet{Erb2014} sample comprises 36 galaxies selected as \lya\ emitters in narrow-band photometry with $-21 \simlt \MUV \simlt -18$, all these objects had rest-frame optical lines detected in MOSFIRE observations. We note the \citet{Erb2014} sample does not include faint \lya\ emitters ($W \simlt 25$\AA), which may have higher velocity offsets given observed anti-correlations between \lya\ EW and $\DV$ \citep{Hashimoto2013,Shibuya2014a,Erb2014}. Whilst these samples are the largest available to measure a $\MUV - \DV$ correlation, future rest-frame optical follow-up of large samples of galaxies with detected \lya\ emission \citep[e.g. from the HETDEX and MUSE-Wide spectroscopic surveys,][]{Song2014,Herenz2017} will provide more complete information about the relationships between \lya\ radiative transfer in the ISM and galaxy properties.

As described above, it is difficult to directly compare galaxies at fixed UV magnitude across cosmic time, so we map UV magnitudes to halo mass. To first order the depth of a halo's gravitational potential well is the dominant influence on galaxy properties independent of redshift \citep{Behroozi2013,Mason2015a,Moster2017}. We assume no redshift evolution between halo mass and velocity offset. We convert UV magnitude to halo mass using the successful model derived by \citet{Mason2015a} which assumes the SFR is proportional to the halo mass assembly rate at a given halo mass and redshift, and is consistent with $\MUV - M_h$ measurements from clustering at $z\sim7$ \citep{Barone-Nugent2014,Harikane2015,Harikane2017}. We add a 0.2 mag uncertainty to the UV magnitude measurements to account for scatter in the UV magnitude -- halo mass relation \citep[e.g.,][]{Finkelstein2015b} and use the propagated uncertainties in halo mass in the $\DV - M_h$. 

In the right panel of Figure~\ref{fig:DV_Muv_lit} we plot the literature $\DV$ measurements as a function of the estimated halo masses. Due to the uncertainties in mapping from UV magnitude to halo mass for very bright galaxies at $z\simlt4$, which may be significantly more starbursty than average, we discard the $z\sim2$ galaxies with $\MUV < -21$ from further analysis. Likewise, we exclude from this inference the galaxies at $z\sim7$ with $\MUV<-22$, deferring their analysis to a later paper \citep{Mason2018}.

When we transform to halo mass the high redshift literature points clearly lie within the low redshift data space. This suggests halo mass is a useful approximately redshift independent indicator of \lya\ escape routes. We note gravitationally lensed objects at intermediate redshifts suggest these trends hold at low mass/luminosity \citep[e.g., a lensed $\MUV=-17$ galaxy at $z\sim3$ was recently observed with a \lya\ velocity offset of 51 km s$^{-1}$,][]{Vanzella2016a}. Further studies, using NIRSpec on JWST, will be able to investigate these trends at high redshifts.

The distribution is well-described by a log-normal distribution with a peak which increases with increasing luminosity, and approximately constant variance:
\BE \label{eqn:pDV_Mh}
p(\DV \,|\, M_h) = \frac{\exp{\left[-\frac{\left(\log_{10} \DV - V(M_h)\right)^2}{2\sigma_v^2}\right]}}{\DV \ln{10} \sqrt{2\pi}\sigma_v } 
\EE
where $V$ is a linear relation corresponding to the most likely $\log_{10}(\DV)$ at a given halo mass:
\BE \label{eqn:DVfromMh}
V(M_h) = m \log_{10}\left(\frac{M_h}{1.55\times10^{12}\msun}\right) + c
\EE
To find the parameters $m, c$ and $\sigma_v$ we take Equation~\ref{eqn:pDV_Mh} as the likelihood function and perform a Bayesian inference on the $z\sim2$ galaxies with $\MUV > -21$, with uniform priors on the parameters. The inferred parameters are: $m=0.32\pm0.07$, $c=2.48\pm0.03$ and $\sigma_v=0.24\pm0.02$. We show this relation on Figure~\ref{fig:DV_Muv_lit}.

We can obtain an approximate relation between velocity offset, UV magnitude and redshift by approximating the \citet{Mason2015a} UV magnitude - halo mass relation as broken linear: $\log_{10} M_h [\msun] \approx \gamma (\MUV + 20.0 + 0.26z) + 11.75$, where $\gamma = -0.3$ for $\MUV \geq -20.0 - 0.26z$, and $\gamma = -0.7$ otherwise. The mean velocity offset in km s$^{-1}$ can then be approximated as:
\BE \label{eqn:DVfromMuv}
\log_{10} \DV(\MUV, z) \approx 0.32\gamma(\MUV + 20.0 + 0.26z) + 2.34
\EE
In this work we sample directly from the distribution in Equation~\ref{eqn:DVfromMh} to calculate velocity offsets directly for simulated halos, including scatter.

In Figure~\ref{fig:DV_Muv_lit} we also plot the circular velocity ($v_c = [10 G M_h H(z)]^{1/3}$) at $z=2$ and $z=7$ for comparison with the observed data. The circular velocities are higher at low redshifts as halos are less dense and more extended, but there is a clear similarity in our derived trend $\DV \sim M_h^{0.32}$ and the circular velocity $v_c \sim M_h^{1/3}$. Investigating these trends with larger samples at low redshifts with dynamical mass measurements \citep[e.g.,][]{Trainor2015,Erb2014} could determine to what extent \lya\ radiative transfer depends on the gravitational potential of the halo.

\subsubsection{Modeling \lya\ line widths}
\label{sec:lyart_ISM_sigma}

The widths of \lya\ lines are also likely dominated by radiative transfer effects which both shifts and broadens the line \citep{Verhamme2006,Verhamme2008,Gronke2016a}. \lya\ velocity dispersions are also observed to be systematically higher than in nebular emission lines which are not resonantly scattered \citep{Trainor2015}.

For simplicity we model the FWHM of the \lya\ lines as equal to the velocity offset of the line, which accounts for the broadening of the lines through scattering and is a good approximation for the observed correlation between \lya\ FWHM and velocity offset \citep[][Verhamme et al. submitted]{Yang2016,Yang2017}.

\subsubsection{EW distribution in an ionized universe}
\label{sec:lyart_pW}

The key observable of \lya\ emission lines at high redshift is their equivalent width (EW or $W$), is a measure of the brightness of the emission line relative to the UV continuum. As \lya\ photons from high redshift galaxies are attenuated by neutral gas in the intervening CGM and IGM we observed only a fraction, $\Tigm$ (the \lya\ \textit{transmission fraction}) of the emitted EW, i.e. $W_\textrm{obs} = W_\textrm{em} \times \Tigm$, where $W_\textrm{em}$ is the emitted equivalent width without any damping due to reionization.

In this work we consider the differential evolution of \lya\ equivalent widths between $z\sim6$ and $z\sim7$, and assume the distribution of equivalent widths changes \textit{only} because of the increasing neutrality of the IGM due to reionization. This is likely a simplification, as trends at lower redshifts show increasing EW with redshift as dust decreases in galaxies \citep{Hayes2011}, but the time between $z\sim6$ and $z\sim7$ is short ($<200$ Myr). If the underlying EW distribution does evolve significantly during that time it will likely be to increase the emitted EW \citep[due to decreasing dust,][]{Hayes2011}, thus the reduction due to reionization would need to be greater to match the observed EW distribution at $z\sim7$ \citep{Dijkstra2011}. \citet{Papovich2011}suggests there may be an increase in gas reservoirs with increasing redshifts due to rapid accretion rates, which could also reduce the emission of strong \lya.

Thus observed equivalent widths at $z\sim7$ are $W_7 = W_6 \times \Tigm_{,7}/\Tigm_{,6}$, where $\Tigm_{,z}$ is the transmission fraction of \lya\ emission for a single object at redshift $z$. In Section~\ref{sec:lyart_IGM} below we describe the calculation of transmission fractions along thousands of lines-of-sight using state-of-the-art cosmological reionization simulations.

A key input to the model then is the $z\sim6$ distribution of EW as a function of galaxy properties. \lya\ EWs for UV continuum-selected galaxies have an observed equivalent width distribution with a peak at zero and some tail to high EW - which is usually parameterized as an exponential function \citep{Dijkstra2012}, log-normal \citep{Schenker2014} or truncated normal distribution plus a delta function \citep{Treu2012}. At $z\simlt2$, where large samples exist, including the local `Green Peas', \lya\ EW is observed to anti-correlate strongly with UV luminosity \citep{Shapley2003,Stark2011,Hashimoto2013} SFR \citep{Verhamme2008}, \HI\ covering fraction \citep{Shibuya2014a} and \lya\ escape fraction \citep{Yang2017}, all indicating \lya\ photons are significantly absorbed by neutral hydrogen gas and dust inside the ISM of massive, highly star-forming galaxies \citep[e.g.][]{Verhamme2008,Erb2014,Yang2017}. At high redshift, the \lya\ EW distribution is usually parameterized as a conditional probability of $p(W \,|\, \MUV)$ \citep{Treu2012,Dijkstra2012}, though dependence on UV spectral slope $\beta$ has also been considered \citep{Schenker2014}.

We take the $z\sim6$ EW distribution from \citet{DeBarros2017} and Pentericci et al. (2018, in preparation) from a Large Program with VLT/FORS2. This sample contains 127 objects, with UV magnitudes between $-22.5 \simlt \MUV \simlt -17.5$, of which 63\% have \lya\ detections. We parameterize it as an exponential distribution plus a delta function:
\BEA \label{eqn:pW_Treu}
p_6(W \,|\, \MUV) &=& \frac{A(\MUV)}{W_c(\MUV)}e^{-\frac{W}
{W_c(\MUV)}} H(W) \nonumber \\
&&+ \left[1 - A(\MUV) \right]\delta(W)
\EEA
$A$ and $W_c$ account for the fraction of non-emitters, and for the anti-correlation of EW with $\MUV$. $H(W)$ is the Heaviside step function and $\delta(W)$ is a Dirac delta function. $A$ implicitly includes contamination by low redshift interlopers in the photometric selection \citep[the interloper fraction is $\leq29\%$ for this sample assuming all non-detections were low redshift contaminants,][]{DeBarros2017}, i.e. we do not distinguish between non-emitters at $z\sim6$ and low redshift contaminants when accounting for non-detections in fitting the parameters (see below). Within our framework this means we assume a similarly small total interloper and non-emitter fraction at $z\sim7$. Recent work by \citet{Vulcani2017} supports this assumption: they found comparably low contamination fractions at $z\sim6$ and $z\sim7$ in an evaluation of photometric selections.

To find these parameters we divided the sample into three bins: $\MUV \leq -21$; $-21 < \MUV < -20$; and $\MUV \geq -20$. We used Equation~\ref{eqn:pW_Treu} as a likelihood ($p_6(W \,|\, A, W_c)$) and performed a Bayesian inference to infer $A$ and $W_c$ for each bin, similar to the methods of \citet{Oyarzun2017}, using uniform priors with $0 < A < 1$ and $0 < W_c < 100$. In the inference we fully account for the the non-detections of \lya\ (using $p_6(W < W_\textrm{lim} \,|\, A, W_c)$ as the likelihood given an EW limit $W_\mathrm{lim}$, in the same way as described in more detail in Section~\ref{sec:bayes} below). The uncertainties and EW limits calculated by \citet{DeBarros2017} are obtained using simulations which fully account for incompleteness and wavelength sensitivity. We note that in this framework the EW limits are a conservative minimum which could be measured over the entire wavelength range (see also Section~\ref{sec:results_current}), future work could incorporate the full wavelength-dependent line flux sensitivity. To allow these parameters to smoothly vary with magnitude between $-21 < \MUV < -20$ we use a hyperbolic tangent function to connect our inferred parameters, without extrapolating beyond the range of the data. We find $A = 0.65 + 0.1\tanh{\left[3(\MUV + 20.75)\right]}$ and $W_c = 31 + 12\tanh{\left[4(\MUV + 20.25)\right]}$\AA\ from fitting to the data. $A$ and $W_c$ vary smoothly with magnitude.

We choose this exponential parameterization of the data because is gives a good description of the data and is easy to treat analytically, and has previously been shown to be an excellent fit to \lya\ EW PDFs \citep[e.g.,][]{Oyarzun2017}. We do not include uncertainties in these parameters and we note the parameterization of $p_6(W \,|\, \MUV)$ is fairly arbitrary but does not qualitatively affect \lya\ modeling during the EoR \citep{Treu2012,Gronke2015b}. Indeed we get the same results, within the uncertainties, if we use the $p_6(W)$ truncated Gaussian distribution from \citet{Treu2012} based on the sample presented in \citet{Stark2011}.

\begin{figure}[t] 
\includegraphics[width=0.49\textwidth]{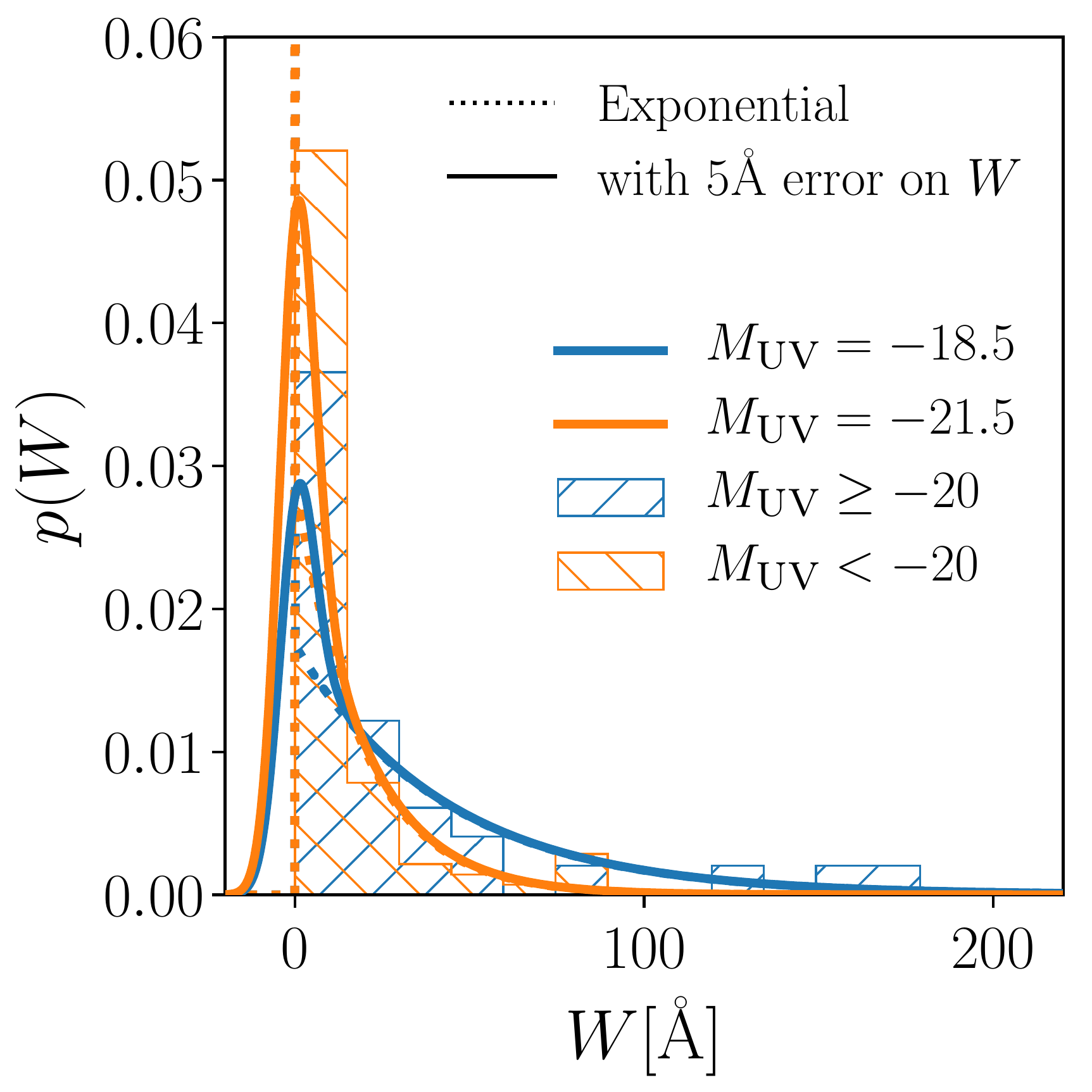}
\caption{$z\sim6$ \lya\ equivalent width distributions for Lyman Break galaxies given by Equation~\ref{eqn:pW_Treu}. The dotted lines show the true distribution. For better comparison with the data, we show the PDFs convolved with a 5\AA\ typical measurement error on $W$ as solid lines. We plot the PDFs for two values of UV magnitude: $\MUV = -18.5, -21.5$ (blue, orange) . UV faint objects tend to have higher EW and a higher duty cycle of \lya\ emission, whereas UV bright galaxies are less likely to emit \lya\ and have lower EWs. We also plot the observed EW from \citet{DeBarros2017} and Pentericci et al. (2018, in preparation) in UV bright (orange) and UV faint (blue) bins. In these histograms we plot all upper limits at $EW=0$, though note we fully account for upper limits in fitting the EW distribution and the reionization inferences (see Equation~\ref{eqn:like_limits}).}
\label{fig:pW_intr}
\end{figure}

Example PDFs given by Equation~\ref{eqn:pW_Treu} are plotted in Figure~\ref{fig:pW_intr} for two values of $\MUV$. We show both the intrinsic PDF and the distribution convolved with a 5\AA\ typical measurement error which introduces at `bump' around $W=0$ where the underlying distribution is a delta function. We also show histograms of the EW observations of \citet{DeBarros2017} and Pentericci et al. (2018, in preparation) in two bins corresponding to UV bright and faint LBGs. As shown by e.g., \citet{Verhamme2008,Stark2010} and \citet{Oyarzun2017}, \lya\ EW strongly depends on UV magnitude.

\subsection{IGM and CGM \lya\ Radiative Transfer}
\label{sec:lyart_IGM}

A \lya\ emission line is significantly attenuated by the CGM and IGM as its photons redshift into resonance with abundant neutral hydrogen along the line-of-sight. Effectively, for a \lya\ line at $z\simgt6$, all photons emitted blue-ward of the \lya\ resonance (1216\AA) are absorbed by the IGM as even after reionization there is still is a fraction of neutral hydrogen within \HII\ regions \citep{Gunn1965}. Infalling overdense gas around halos can also increase the opacity of the IGM near the \lya\ resonance and onto the red side of the \lya\ line \citep{Santos2004,Dijkstra2007,Laursen2011}.

For simplicity we assume all \lya\ photons emitted below the circular velocity of a halo are absorbed in the CGM, and all redder photons are transmitted \citep{Dijkstra2011,Laursen2011}. This treatment of the CGM may be crude but it enables us to investigate the relative difference between observations at $z\sim6$ and $z\sim7$ assuming any evidence of a difference is driven by reionization. Future work could incorporate more complex modeling of CGM absorption \citep[e.g.,][]{Kakiichi2017}. Figure~\ref{fig:lineprofile} show example model \lya\ emission lines, where the dotted black lines correspond to the intrinsic line profile after transmission through the ISM and the black solid lines correspond to the lineshape after resonant absorption in the CGM/IGM which absorbs the flux blueward of $v_\mathrm{circ}$.

During reionization, there is an additional opacity to \lya\ caused by the presence of cosmic diffuse neutral hydrogen patches which attenuate the damping wing of the \lya\ line cross-section \citep{MiraldaEscude1998a}. The transmission of \lya\ photons through the reionizing IGM is driven by the global fraction of neutral hydrogen, $\xHI(z)$.

Thus the total opacity to \lya\ due to neutral hydrogen within the IGM is given by:
\BE \label{eqn:tau_IGM}
\tau_\textsc{igm}(z, v) = \tau_\textsc{d}(z, v) + \tau_\mathrm{\HII}(z, v)
\EE
where $\tau_\textsc{d}(z, v)$ is the damping wing optical depth which is only present during the EoR, and $\tau_\mathrm{\HII}(z, v)$ is the optical depth due to resonant absorption within the CGM of galaxies (infalling gas) and any neutral hydrogen within the local \HII\ region of a galaxy. For simplicity, we assume $e^{-\tau_\mathrm{\HII}} = H(v-v_\mathrm{circ})$ at both $z\sim6$ and $z\sim7$.

In this model, we assume the universe is fully ionized at $z\sim6$, thus the damping wing opacity only becomes important at $z>6$. This may not be exactly the case, but current constraints on $\xHI$ at $z\sim6$ suggest the neutral fraction is low \citep[$\xHI < 0.1$,][]{McGreer2014} so the reionization effect on \lya\ emission will be small. 

To obtain the damping wing optical depths $\tau_\textsc{d}(z=7, v)$ requires a model of the IGM topology during reionization. Whilst observation papers of \lya\ emission with reionization inferences have used simple `patchy' or `smooth' IGM topologies \citep{Treu2012,Treu2013,Pentericci2014,Tilvi2014}, for this work, we consider realistic reionization topologies from state-of-the-art theoretical modeling. We obtain \lya\ damping optical depths from the public Evolution of 21cm Structure (EoS) suite of reionization simulations described by \citet{Mesinger2014,Mesinger2016}\footnote{\url{http://homepage.sns.it/mesinger/EOS.html}}. 

Due to the strong clustering of the first galaxies spatial fluctuations in the IGM neutral fraction during reionization existed on scales of tens of Mpcs. Accurately modeling these fluctuations and the growth of ionized \HII\ bubbles in the IGM requires cosmological simulations at least 100 Mpc in size \citep{Trac2011,Mesinger2014}. The EoS reionization simulations use \textsc{21cmfastv2} \citep{Sobacchi2014} where inhomogeneous recombinations and ionizations in the IGM are treated at a sub-grid level on a density field in a box with sides 1.6 Gpc with a resolution $1024^3$. The simulations produce $\xHI$ maps at different redshifts and superimpose them on the $z\sim7$ halo field to produce cubes of the $z\sim7$ IGM for a range of neutral fractions. For the bulk of reionization, this is analogous to changing the ionization efficiency at a fixed redshift \citep[e.g.,][]{McQuinn2007a,Mesinger2008a}.

The timeline and topology of reionization is determined by the mass of galaxies which dominate reionization and the ionization efficiency, $\zeta \propto f_\mathrm{esc} \times f_\star$, where $f_\mathrm{esc}$ is the fraction of ionizing photons which escape galaxies into the IGM, and $f_\star$ is the stellar mass fraction in galaxies. As both of these parameters are expected to scale with halo mass in complementary ways \citep[i.e. low mass halos host galaxies with a low stellar mass fraction and high escape fraction, e.g.,][]{Kimm2016,Trebitsch2017}, over the relevant range of halo masses which host galaxies which dominate reionization \citep[$M_h \simlt 10^{11} \msun$, e.g.,][]{Kimm2016}, $\zeta$ is assumed to be constant in the EoS simulations. The simulations use a free parameter which sets the minimum mass of halos capable of hosting star formation, and then adjust $\zeta$ to produce a Thompson scattering optical depth to the CMB consistent with \citet{PlanckCollaboration2015}.

We use the fiducial `Faint Galaxies' run which corresponds the primary drivers of reionization being low mass star-forming galaxies with an atomic cooling threshold of $T_\mathrm{vir} \simgt 10^4$ K, with $\zeta=20$, producing IGM morphologies characterized by small HII regions. Whilst the EoS simulations have another run, `Bright Galaxies', where reionization is driven by more massive galaxies, producing larger HII regions, it has been shown that information about the reionization morphology is mostly smeared out when using galaxies spread in redshift \citep[$\Delta z \simgt 0.1$ bin, e.g.,][though with large spectroscopic samples, $\Delta z \sim 0.01$, the sensitivity does increase]{Sobacchi2015a}, as is the case for our sample (see Section~\ref{sec:results_current}), so we do not expect a significant change in our results if we were to use an alternative run. Indeed, \citet{Mason2018} uses both simulation runs but shows that the transmission of \lya\ from galaxies $M_h \sim 10^{10} - 10^{12}$ is relatively independent of the reionization morphology. Similarly, \citet{Greig2016b} show QSO damping wing effects are not particularly sensitive to the reionization morphology.

Halos are located in the same density field as the IGM simulation. We ignore absorption from Damped \lya\ Absorbers (DLAs) inside the cosmic \HII\ regions \citep{Bolton2013} which has been shown to have a minor impact on the \lya\ fraction when self-shielding is calculated more accurately \citep{Mesinger2014}. We refer the reader to \citep{Mesinger2016} for more details of the simulation. For this work we focus on $z\sim7$, where large samples of LBGs have spectroscopic follow-up \citep{Pentericci2014,Schmidt2016}, but it is easy to extend the work to any other redshift.

\begin{figure}[t] 
\includegraphics[width=0.49\textwidth]{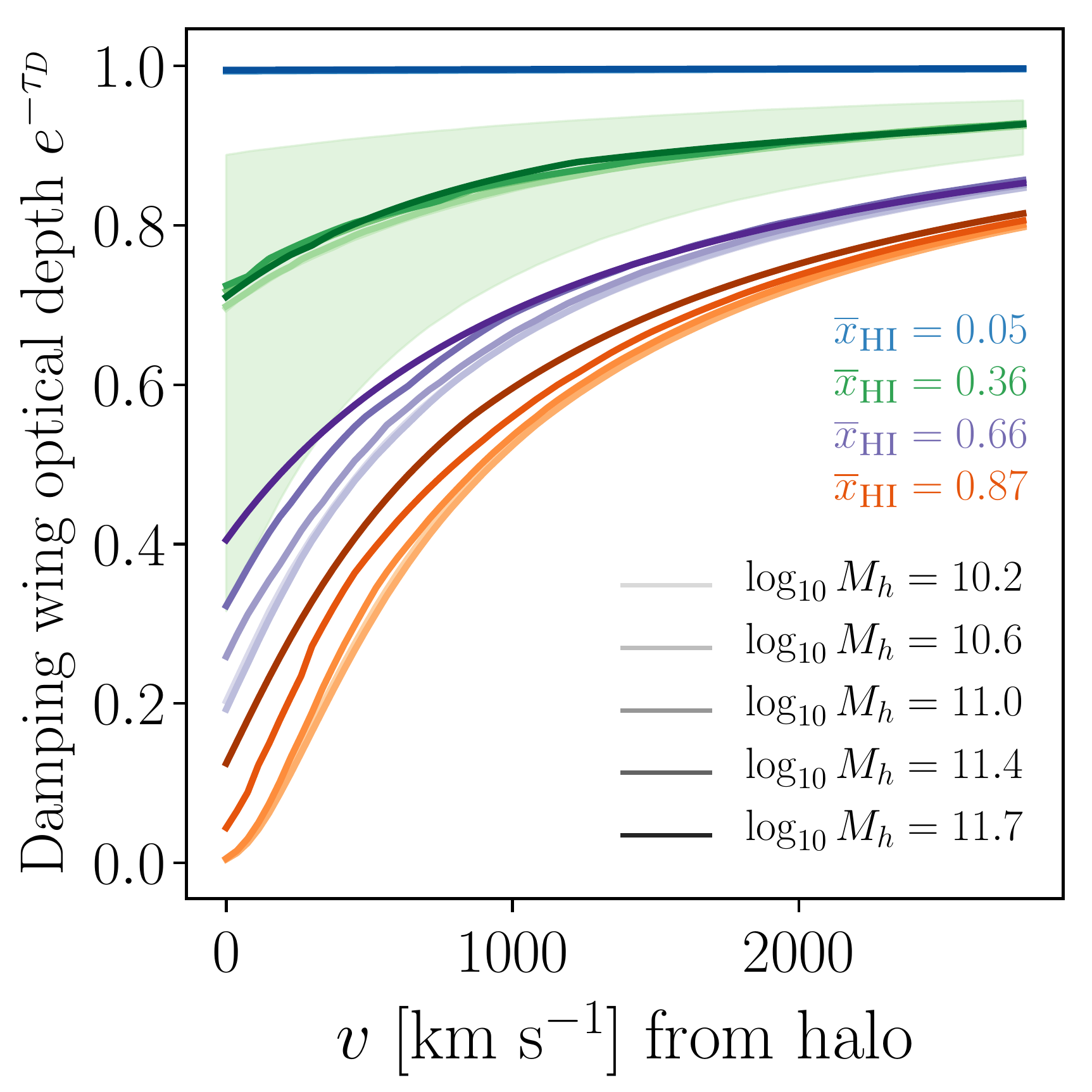}
\caption{Median \lya\ IGM damping wing optical depths due to cosmic \HI\ patches during reionization as a function of velocity offset from the center of the source halos. We plot optical depths for 5 different mass halos (indicated by tone of the line - where darkest lines are the highest mass halos) and for 4 volume-averaged neutral fractions $\xHI$ (indicated by color). We plot the median optical depth for each halo from the 1000s of simulated sightlines. For $\xHI=0.36$ we plot the 1$\sigma$ confidence region for the optical depths from all the sightlines to the $\log_{10}M_h = 10.2$ halos as a shaded area, showing the large variation across sightlines.}
\label{fig:optdepth_Mh}
\end{figure}

We take 1000s of sightlines emanating from halos with masses $\sim 10^{10-12} \msun$ \citep[comparable to typical $z\simgt5$ halo masses for $-22 \simlt \MUV \simlt -18$ galaxies,][]{Barone-Nugent2014,Harikane2015,Harikane2017} and compute the damping wing optical depth, $\tau_\textsc{d}$, for \lya\ emission as a function of velocity offset from the systemic redshift of the source halos in boxes with a range of global neutral fractions. Median values of $\exp{[-\tau_\textsc{d}]}$ along $\sim50$ (to the rarest high mass halos) to $\simgt4000$ (to typical $10^{10.5} \msun$ halos) sightlines are plotted in Figure~\ref{fig:optdepth_Mh} for a range of halo masses and $\xHI$. The optical depths are smooth functions of velocity and clearly damp \lya\ more effectively for higher $\xHI$. In general, higher mass halos have lower optical depths to \lya\ as their large bias means they are more likely to live in the centers of large \HII\ regions, relatively more distant from the cosmic \HI\ patches which produce the damping wing absorption during the EoR.

For a given sightline, the final fraction of \lya\ photons emitted by a galaxy in a halo with mass $M_h$ which are transmitted through the IGM, $\Tigm$, is given by:
\BE \label{eqn:Tfromtau}
\Tigm(\xHI, M_h, \DV) = \int_0^\infty \mathrm{d}v \; J_\alpha(\DV, M_h, v) e^{-\tau_\textsc{igm}(\xHI, M_h, v)}
\EE
where $\DV$ is the velocity offset of the \lya\ line center from the systemic redshift of the source galaxy (which depends on the galaxy's ISM, as described in Section~\ref{sec:lyart_ISM}) and $J_\alpha(\DV, M_h, v)$ is the line profile of \lya\ photons escaping from the galaxy as function of velocity $v$. 

As any photons emitted bluer than the halo circular velocity will be resonantly absorbed by intervening neutral hydrogen \citep{Gunn1965,Dijkstra2007,Zheng2010,Laursen2011,Schroeder2013}, $J_\alpha$ is described as:
\BE \label{eqn:Jalpha}
J_\alpha(\DV, M_h, v) \propto 
\begin{cases}
    \frac{1}{\sqrt{2\pi}\sigma_\alpha}e^{-\frac{(v - \Delta v)^2}{2\sigma_\alpha^2}}& \text{if } v \geq v_\mathrm{circ}(M_h)\\
    0              & \text{otherwise}
\end{cases}
\EE
If $J_\alpha$ is normalized $\Tigm_{,6} = 1$, as we assume $\tau_\textsc{d} = 0$ at $z\sim6$. Compared to the intrinsic emitted line $\Tigm_{,6}$ can be very low \citep{Dijkstra2007,Zheng2010,Laursen2011}. For ease of notation we refer to the differential transmission at $z\sim7$, $\Tigm_{,7}/\Tigm_{,6}$, as $\Tigm$.

\begin{figure}[t] 
\includegraphics[width=0.49\textwidth]{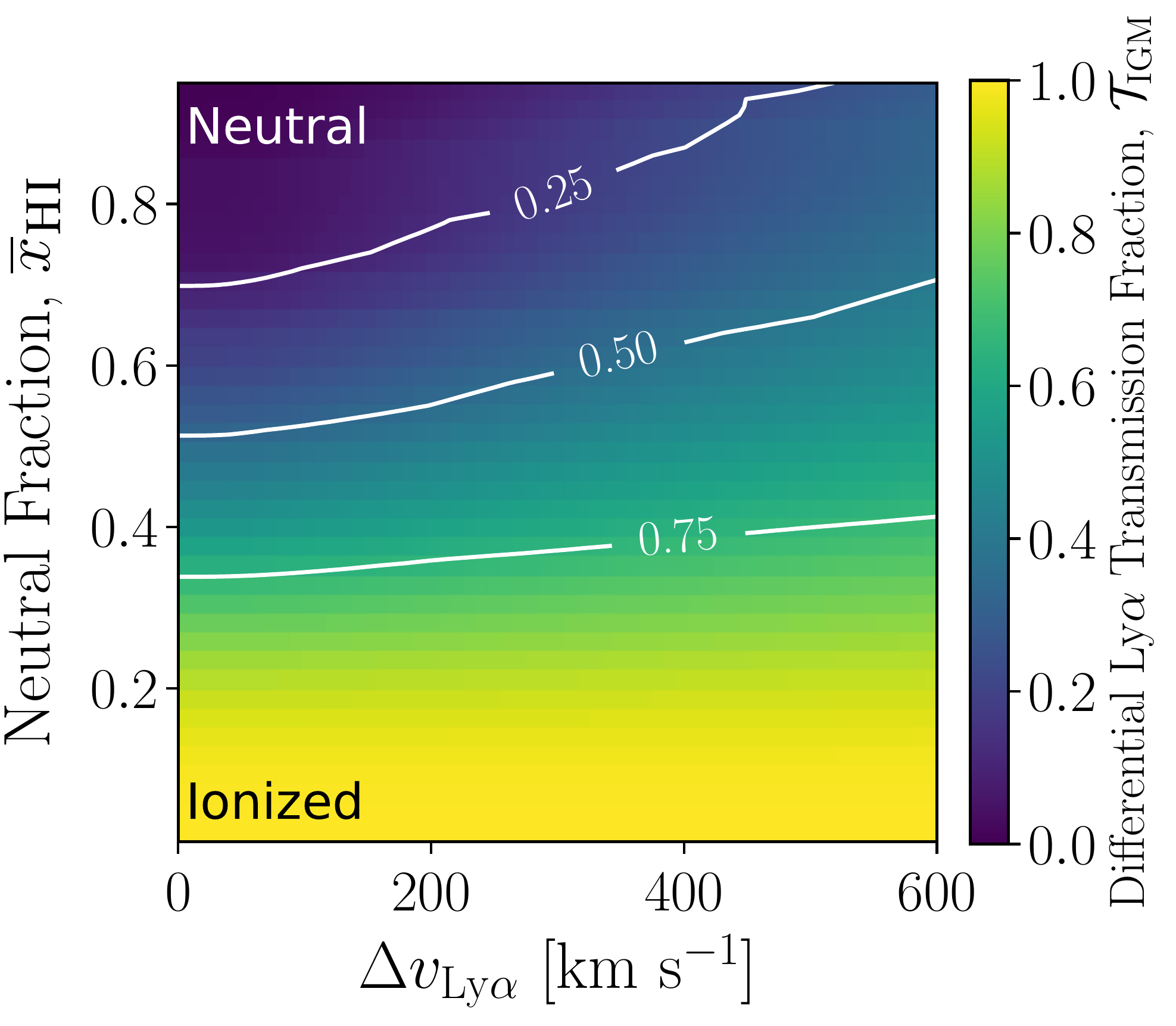}
\caption{Median fraction of \lya\ photons transmitted through the IGM, $\Tigm$, as a function of $\xHI$ and $\DV$ computed with Equation~\ref{eqn:Tfromtau} from $\sim$5000 sightlines to halos with mass $10^{10} \msun$, assuming $\Tigm_{,6} = 1$. Contours show transmission fractions of 25\%, 50\% and 75\%. In a predominantly neutral universe \lya\ photons have higher probability of escape through predominately ionized IGM and if emitted at high velocity offsets from their originating galaxies.}
\label{fig:T}
\end{figure}

Example intrinsic and transmitted emission lines are plotted in Figure~\ref{fig:lineprofile}. Sightline median values of $\Tigm(\xHI, \DV)$ at fixed halo mass are plotted in Figure~\ref{fig:T}. As expected, as the neutral fraction increases the transmission fraction of \lya\ decreases smoothly. Whilst at low neutral fractions the velocity offset of \lya\ has little impact, in a predominantly neutral universe, $(\xHI \simgt 0.6)$ lines are more easily transmitted if they were emitted at high velocity offset.

\begin{figure}[t] 
\includegraphics[width=0.49\textwidth]{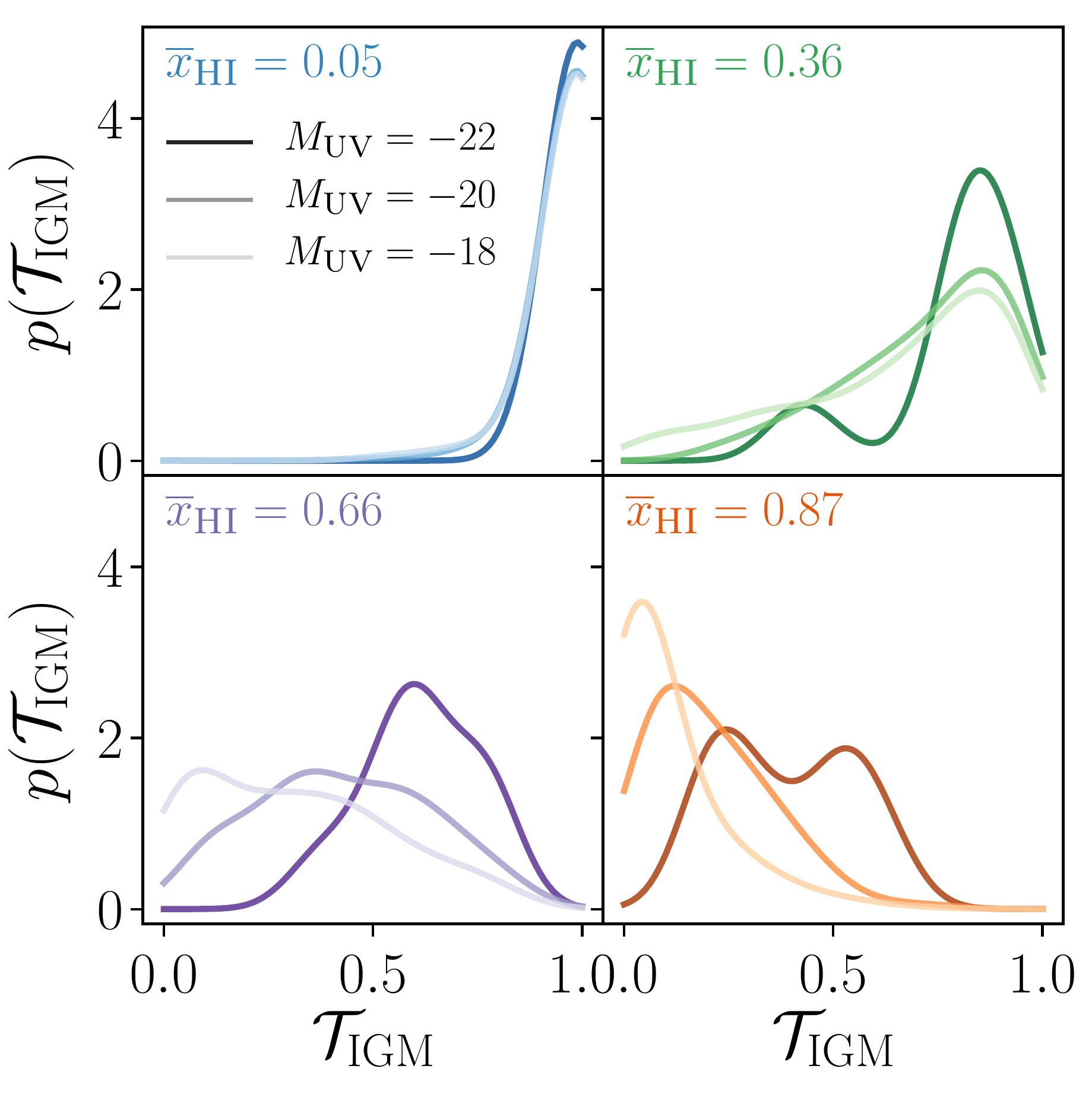}
\caption{Distributions of differential \lya\ transmission fractions $\Tigm$ at $z\sim7$ for simulated galaxies of different UV luminosities (UV bright = darkest lines), for a range of IGM neutral fractions $\xHI$. As described in Section~\ref{sec:lyart_IGM} this is the ratio of \lya\ transmission at $z\sim7$ to that at $z\sim6$ where there is already significant absorption within the ionized IGM \citep{Dijkstra2007,Zheng2010,Laursen2011}. The transmission fractions evolve smoothly with the neutral fraction, though the evolution is more gradual for UV bright galaxies.}
\label{fig:T_Muv}
\end{figure}

In Figure~\ref{fig:T_Muv} we plot probability distribution functions for $\Tigm$ for three different values of $\MUV$, where we have transformed from halo mass to $\MUV$ using the \citet{Mason2015a} LF model as above and drawn $\Delta v$ values for halos using the distribution in Equation~\ref{eqn:DVfromMh}. The transmission distributions evolve smoothly with neutral fraction and UV magnitude. Transmission of \lya\ evolves more slowly for the brightest galaxies, due to a combination of their increased velocity offsets and their locations in the most overdense regions, far from the cosmic \HI\ regions which cause the damping wing absorption.

Galaxies in high mass halos ($M_h>10^{12} \msun$, corresponding to approximately $\MUV<-22$) require special attention. First, they are rare and lines of sights to such objects in the simulations are not well-sampled leading to large statistical errors. Second, the correlation between $\MUV$ and $M_h$ is particularly uncertain in this regime. Third, such bright galaxies have been observed to buck the trend in the declining \lya\ emission fraction at $z\simgt7$ at $z>7.5$ \citep{Curtis-Lake2012,Stark2017}. For these reasons, they require special attention, especially because they are prime targets for detailed spectroscopic follow-up. Since they are intrinsically rare, they would contribute negligibly to the analysis presented in this paper. Thus, we leave their analysis for future work \citep{Mason2018} and exclude them from the sample considered here.

\newpage
\section{Bayesian Inference}
\label{sec:bayes}

\begin{figure}[t] 
\includegraphics[width=0.49\textwidth]{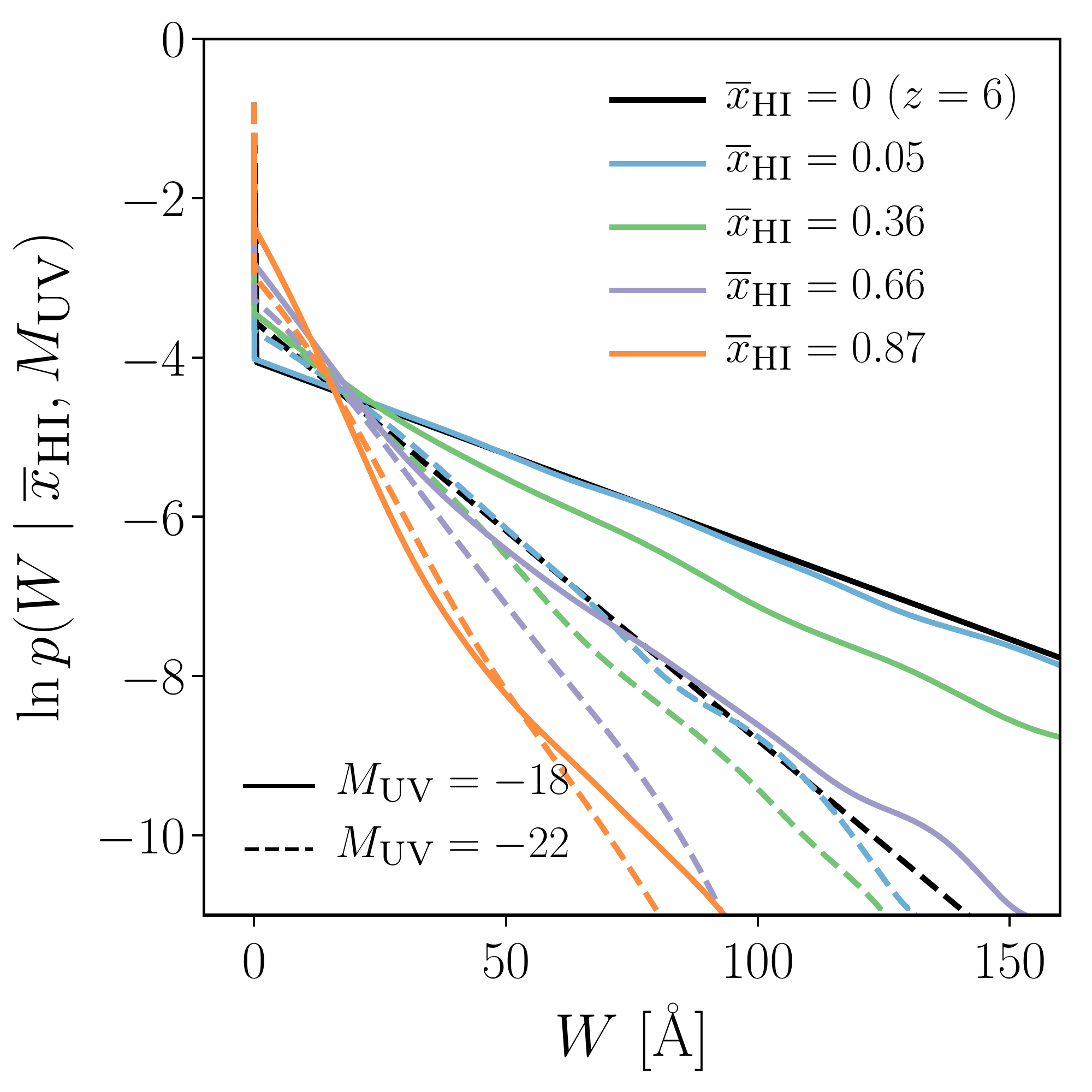}
\caption{Simulated observed distribution of \lya\ equivalent widths (the likelihoods for our model) for a range of neutral fractions (colors), for faint (solid line) and bright (dashed line) UV magnitudes. The intrinsic distributions (Equation~\ref{eqn:pW_Treu}) are shown as black lines. The EW distribution evolves significantly for the UV faint galaxies with increasing $\xHI$, whilst the distribution for UV bright galaxies evolves more slowly.}
\label{fig:like}
\end{figure}

Bayes' Theorem enables us to infer the posterior distribution of model parameters of interest, $\theta$ given our observed data $Y$ from the likelihood of obtaining the data given our model and our prior information of the model parameters. The posterior probability of $\theta$ is written as:
\BE \label{eqn:bayes}
p(\theta \,|\, Y) = \frac{p(Y \,|\, \theta)\, p(\theta)}{p(Y)} 
\EE
where $p(Y \,|\, \theta)$ is the likelihood function, $p(\theta)$ is the prior, and $p(Y)$ is the Bayesian Evidence which normalizes the posterior.

We want to obtain the posterior distribution of the volume averaged fraction of neutral hydrogen, $\xHI$, a global IGM property, given the observed data: measurements of \lya\ equivalent widths $W$ and galaxy rest-frame UV absolute magnitudes $\MUV$. As described in Section~\ref{sec:lyart} we model both IGM and ISM effects on \lya\ transmission and produce forward models of the observed \lya\ equivalent widths for galaxies of a given UV magnitude.

Using Bayes' Theorem we can write the posterior probability for $\xHI$ inferred from one observation in the absence of noise as:
\BE \label{eqn:post}
p(\xHI \,|\, W, \MUV) \propto p(W \,|\, \xHI, \MUV)\, p(\xHI) 
\EE
where $p(W \,|\, \xHI, \MUV)$ is the likelihood of observing a \lya\ equivalent width given our forward model of the ISM and IGM, and $p(\xHI)$ is the prior on the neutral fraction which we assume is uniform between 0 and 1.

Usually, the likelihood function is obtained from a model with an analytic form - e.g. a normal distribution, however, due to including simulated IGM cubes, our model is complex and does not have an analytic parameterization. We therefore generate the likelihood by sampling $10^6$ realizations of galaxies in our model at a given $(\xHI, \MUV)$ and then perform a Kernel Density Estimation \citep{Rosenblatt1956,Parzen1962} to fit a smooth probability density function to the sampled distribution. Examples of the likelihood function are shown in Figure~\ref{fig:like}. Generation of the likelihoods is described in more detail below in Section~\ref{sec:bayes_like}.

In reality, our observations will always have measurement uncertainties, and some observations can only place an upper limit on a measurement, given a noise level. When we include noise, our likelihood for measuring an equivalent width $W_i$ with Gaussian noise level $\sigma_i$ becomes:
\BE \label{eqn:like_noise}
p(W_i \,|\, \xHI, \MUV) = \int_0^\infty \mathrm{d}W \; \frac{e^{-\frac{(W - W_i)^2}{2\sigma_i^2}}}{\sqrt{2\pi}\sigma_i} p(W \,|\, \xHI, \MUV)
\EE
and the likelihood for upper limits, $W_i < \mathcal{W}$ is given by:
\BEA \label{eqn:like_limits}
p(W_i < \mathcal{W} \,|\, \xHI, \MUV) &=& \int_{-\infty}^\mathcal{W} \mathrm{d}W \; p(W_i \,|\, \xHI, \MUV) \\
	&=& \int_0^\infty \mathrm{d}W \; \frac{1}{2} \erfc{\left(\frac{W - \mathcal{W}}{\sqrt{2}\sigma_i} \right)}  \nonumber \\  &&\times p(W \,|\, \xHI, \MUV) \nonumber 
\EEA
where $\erfc(x)$ is the complementary error function for $x$. 

In this work we consider samples with good redshift completeness (i.e. the probability of a \lya\ line falling within the observable range is close to one, see Section~\ref{sec:results_current}). Thus, this inference framework does not include information about redshift in the likelihood, this is left for future work.

We can combine the inference from a set of independent observations (i.e. individual galaxies) by simply multiplying the posteriors:
\BE \label{eqn:like1}
p(\xHI \,|\, \{W, \MUV\}) \propto \prod_{i}^{N_\textrm{gals}} p(W_i \,|\, \xHI, M_{\textsc{uv},i}) \, p(\xHI) 
\EE
%

\subsection{Generating the likelihood}
\label{sec:bayes_like}

Our observed data are a set of \lya\ equivalent widths (and limits) and absolute magnitudes from galaxies at a given redshift: $\{W, \MUV \}$. Due to the complexity of the IGM topology, there is no simple analytic model to express the likelihood of obtaining these data given a neutral fraction $\xHI$. Thus we use our model to generate large samples of mock observations which provide a non-analytic likelihood.

We take IGM simulations with global neutral fractions $0.01 \leq \xHI \leq 0.95$ ($\Delta \xHI \sim 0.02$) and a population of halos with masses $10^{10} \simlt M_h [\msun] \simlt 10^{12}$ with $\Delta \log M_h \sim 0.1$. This mass range corresponds to UV magnitudes of $-16 \simgt \MUV \simgt -22$ at $z\sim7$ \citep{Mason2015a}. The likelihood is computed in the following way:

\begin{enumerate}
\item Obtain the \lya\ damping wing optical depths (see Section \ref{sec:lyart_IGM}) along thousands of different sightlines to individual halos in each simulation, to account for the inhomogeneous nature of reionization.
\item For a grid of UV magnitudes $-22 \leq \MUV \leq -17$ we nearest-neighbor match the simulation halo masses with UV magnitudes at $z\sim7$ given by the relation in \citet{Mason2015a} which is consistent with $\MUV - M_h$ measurements from clustering at $z\sim7$ \citep{Barone-Nugent2014,Harikane2015,Harikane2017}. We do not add scatter to this matching, but note the halo mass step in the simulations ($\sim0.13$ dex) is not dissimilar to the scatter in the inferred $\MUV-M_h$ relation for galaxies around $\MUV^\star$ \citep[e.g., $0.3$ dex, ][]{Finkelstein2015b}, so some $\MUV - M_h$ scatter is included. Furthermore we note the optical depth scatter between sightlines for a given halo mass is far greater than the scatter between sightlines between halos of different masses (compare lines and shaded region in Figure~\ref{fig:optdepth_Mh}), thus the $\MUV - M_h$ scatter is sub-dominant.
\item Populate these model galaxies with \lya\ line velocity offsets from the distribution $p(\DV \,|\, M_h)$ as described by Equation~\ref{eqn:DVfromMh}, including the scatter $\sigma_v$, and the \lya\ equivalent widths for an ionized universe (we which assumed to be the same as at $z\sim6)$, $p_6(W_\mathrm{em} \,|\, \MUV)$ described in Section~\ref{sec:lyart_ISM}, creating $10^6$ realizations of a galaxy in each halo.
\item We compute the differential \lya\ transmission fraction, $\Tigm$ with Equation~\ref{eqn:Tfromtau} along sightlines through the IGM to every model galaxy and the observed equivalent width, where $W_\mathrm{obs} = \Tigm \times W_\mathrm{em}$. 
\item The distributions of model observed $W_\mathrm{obs}$ at fixed $(\xHI, \MUV)$ are described by the form:
\BEA \label{eqn:like_KDE}
p(W \,|\, \xHI, \MUV) &=& A(\MUV)f(W, \xHI) H(W) \nonumber \\
						 &&+ \left[1 - A(\MUV) \right]\delta(W)
\EEA
where $f(W, \xHI)$ describes the evolution of the equivalent width distribution as the neutral fraction evolves and is fitted with a Gaussian Kernel Density Estimator \citep{Rosenblatt1956,Parzen1962}, and $A(\MUV)$ denotes the fraction of non-emitters and contaminants as described in Equation~\ref{eqn:pW_Treu} which does not change as the neutral fraction increases ($\Tigm \neq 0$ exactly).
\end{enumerate}

These distributions $p(W \,|\, \xHI, \MUV)$ are the likelihoods for the observed data. Some examples are plotted in Figure~\ref{fig:like}. For increasing neutral fraction the EW distribution becomes steeper, as more \lya\ is damped by cosmic neutral patches. The evolution of $p(W \,|\, \xHI, \MUV)$ is slower for more UV bright (more massive) galaxies because the transmission functions evolve more slowly with increasing neutral fraction (see Section~\ref{sec:lyart_IGM} and Figure~\ref{fig:T_Muv}).

We chose to marginalize out $\DV$ at this stage to ease computation by reducing a degree of freedom, but it is possible to produce the likelihood conditional on $\DV$: $p(W \,|\, \xHI, \MUV, \DV)$. It is then possible to infer $\DV$ for an individual observed galaxy, or, if $\DV$ is already known, recover a narrower posterior on the neutral fraction. 

\section{Results}
\label{sec:results}

In this section we describe the key results and predictions from our model. In Section~\ref{sec:results_model} we show our method can accurately recover the neutral fraction for simulated datasets. We perform inference on current data from \citet{Pentericci2014} in Section~\ref{sec:results_current}. In Section~\ref{sec:results_jwst} we make predictions for future surveys with JWST.

\subsection{Large samples of galaxies can accurately constrain the neutral fraction}
\label{sec:results_model}

\begin{figure}[] 
\includegraphics[width=0.49\textwidth]{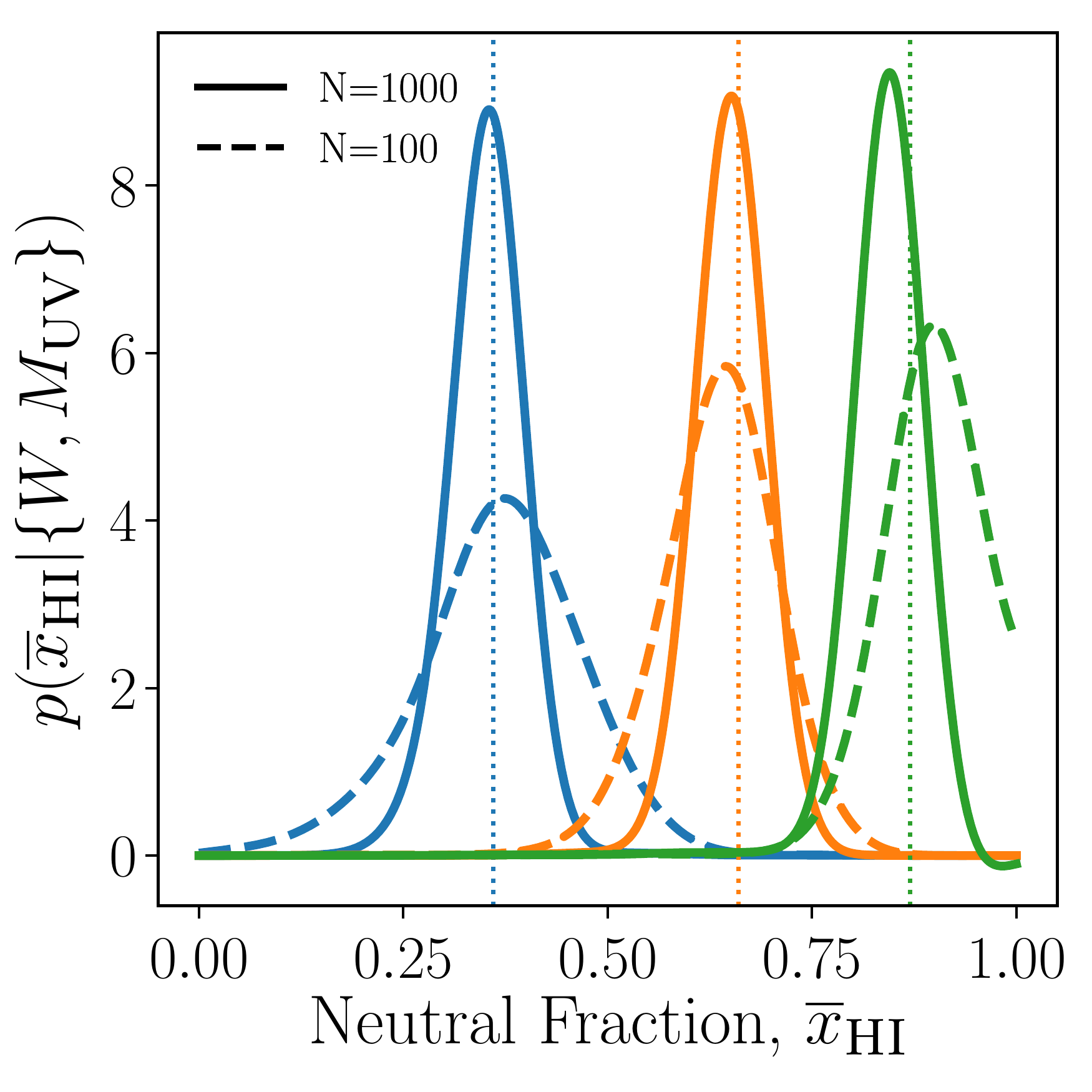}
\caption{Posterior distributions for $\xHI$ from simulated samples of \lya\ detections from 1000 (solid lines) and 100 (dashed lines) galaxies, for a simulation input value of $\xHI=[0.36, 0.66, 0.87]$ (blue, orange, green - the input value is shown by the vertical dotted line). With large samples input neutral fraction is recovered well. With smaller samples, the posterior is wider, but includes the true value within 1$\sigma$ uncertainty.}
\label{fig:posterior_test}
\end{figure}

To test our inference framework we perform simulated surveys of LBG follow-up. We draw a realistic sample of LBGs at $z\sim7$ from the \citet{Mason2015a} UV luminosity function model \citep[which is consistent with all observations, including new deep data from the Hubble Frontier Fields at $z\simgt7$ e.g.,][]{Atek2015a,McLeod2016,Livermore2017,Bouwens2016a}. We populate these galaxies with an EW given by our simulated $p(W \,|\, \xHI, \MUV)$ (see Section~\ref{sec:bayes_like}) for several test values of the neutral fraction.

We assume an apparent magnitude limit of $m_\textsc{ab} = 28.5$, corresponding to $\MUV \sim -18.5$ and a 5$\sigma$ flux limit of $10^{-18}$ erg s$^{-1}$ cm$^{-2}$. We draw samples of 100 and 1000 total galaxies, and perform the inference on the full samples including upper limits.

In Figure~\ref{fig:posterior_test} we plot the resulting posterior distributions for $\xHI$. With large samples we can clearly recover the input neutral fraction well. With small samples the posterior distribution is broader as we sample less of the likelihood, but the posteriors still include the input value within 1$\sigma$.

\newpage
\subsection{Inference from current data}
\label{sec:results_current}

We use the inference framework described above to infer the neutral fraction from current observations. We take the largest published sample of LBGs at $z\sim7$ with spectroscopic follow-up to-date, presented in \citet{Pentericci2014}. These data comprise 68 galaxies spanning UV magnitudes $-22.75 \simlt \MUV \simlt -17.8$ and include 10 intrinsically faint objects gravitationally lensed behind the Bullet Cluster \citep{Bradac2012} as well as observations in deep HST legacy fields \citep{Fontana2010a,Vanzella2011,Ono2012,Schenker2012}. 

\begin{figure}[t] 
\includegraphics[width=0.49\textwidth]{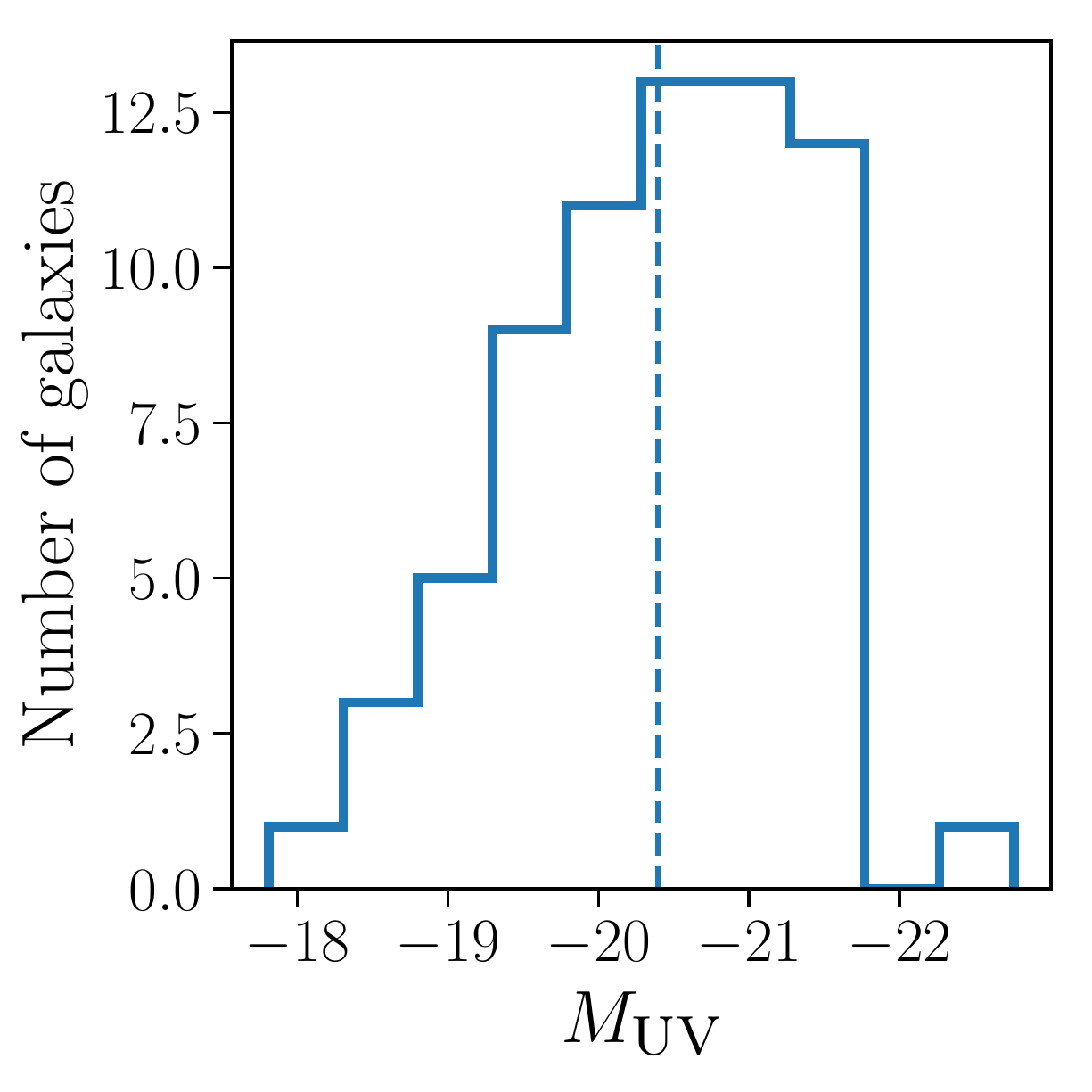}
\caption{UV magnitude distributions for the $z\sim7$ sample used for the inference. We plot the median UV magnitude of the sample as a dashed line ($\MUV = -20.4$)}
\label{fig:hist_pentericci_Muv}
\end{figure}

\begin{figure}[t] 
\includegraphics[width=0.49\textwidth]{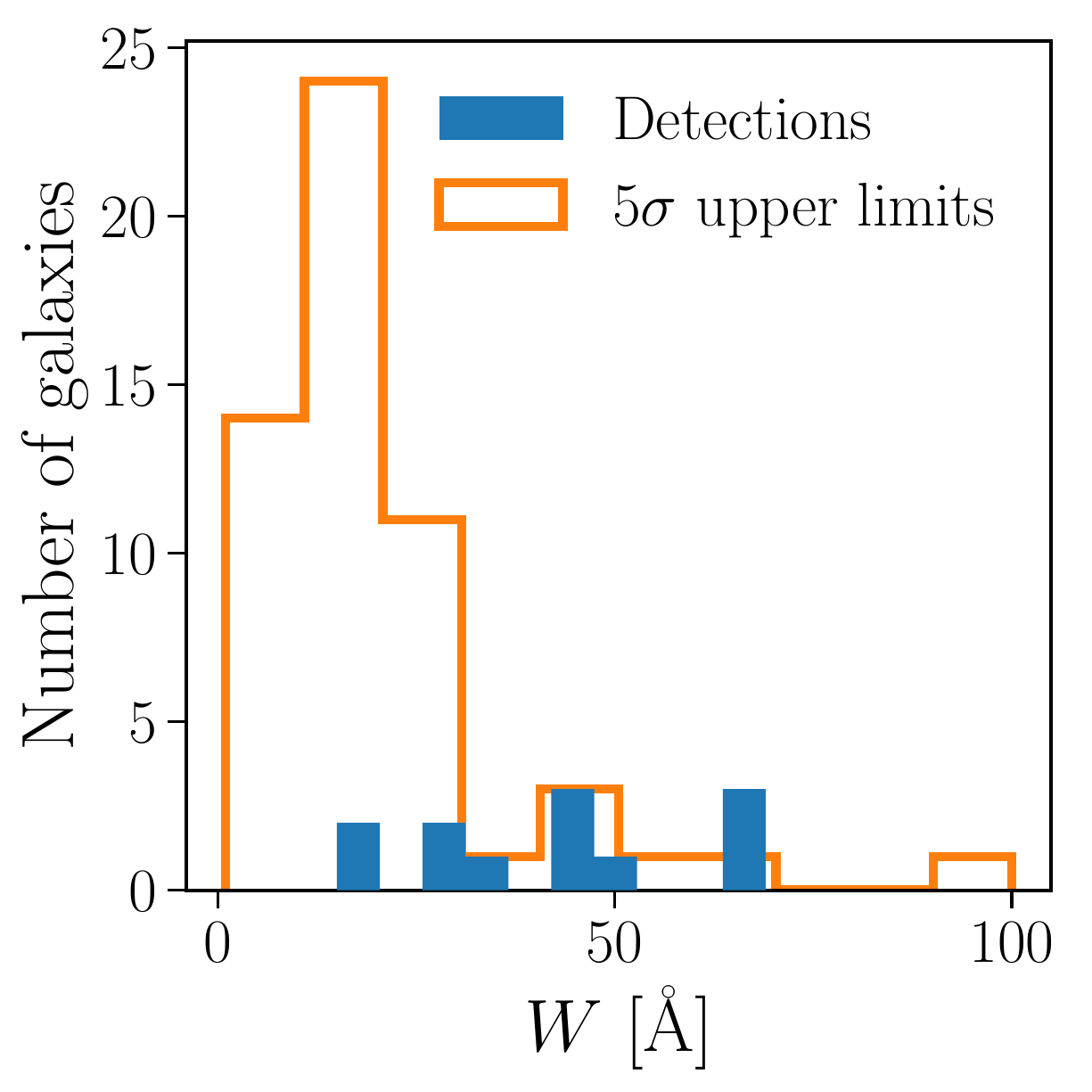}
\caption{EW distribution for the $z\sim7$ sample used for the inference. We show both the \lya\ EW measurements (filled blue) and $5\sigma$ upper limits for the non-detections (orange line).}
\label{fig:hist_pentericci_EW}
\end{figure}

In total, the sample comprises 8 independent lines-of-sight with field areas $\sim50-100$ arcmin$^2$ each. The detections are spread over these fields. \citet{Pentericci2014} quantified the cosmic variance in this sample is very low \citep[$\sim6\%$ uncertainty in the optical depth to \lya, see also][]{Trenti2008}. Of the 68 LBGs 12 \lya\ lines were spectroscopically confirmed. In Figure~\ref{fig:hist_pentericci_Muv} we plot the UV magnitude and in Figure~\ref{fig:hist_pentericci_EW} the EW distributions for this sample. As for the \citet{DeBarros2017} $z\sim6$ sample, the EW limits are obtained by inserting simulated lines of varying flux, FWHM, and redshift into raw data and then trying to recover them. A conservative minimum flux that could be measured over the entire wavelength range is used for the EW limit. Our framework utilizes the fact that the non-detections must be fainter than this conservative limit: fainter lines could be observed, e.g., in regions free of sky lines. Future work could include the wavelength-dependent line flux sensitivity in the likelihood for non-detections (Equation~\ref{eqn:like_limits}).

The majority of targets were $z_{850}$-dropouts selected primarily using a color criteria technique, described in detail in \citet{Grazian2012}, to find targets with a high probability of having redshifts $6.5 \simlt z \simlt 7.5$. The median redshift for this selection function was $z=6.9$ \citep[see Figure 1 in][]{Grazian2012}. For literature targets not directly observed by the \citet{Pentericci2014} group, but included in the sample, they only included $z$-dropouts with colors consistent with the color selection criteria. As noted by \citet{Pentericci2014} the probability of galaxies being outside of the observable range for their setup ($z\sim7.3$) is negligibly, except for the 10 objects in the Bullet Cluster \citep{Bradac2012} where $\sim48\%$ of objects could be above this redshift due to the broad J filter used for selection of that sample \citep{Hall2012}. To test the impact of these few objects potentially being at higher redshifts we ran the inference excluding the Bullet Cluster and found it does not significantly impact the results.

\begin{figure}[t] 
\includegraphics[width=0.49\textwidth]{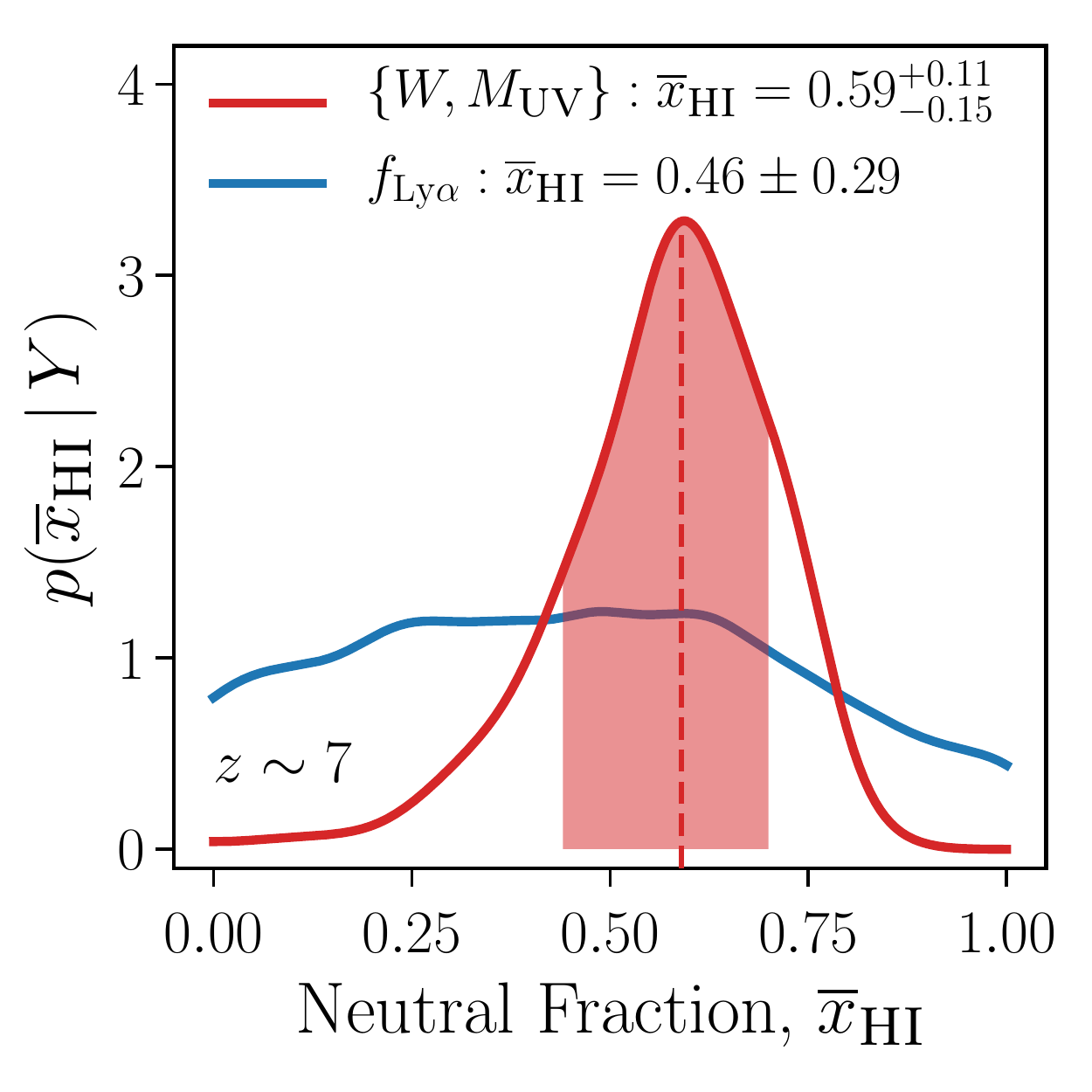}
\caption{Posterior distribution for $\xHI$ from the dataset of 68 galaxies at $z\sim7$ (including 12 with detected \lya\ emission) from \citet{Pentericci2014}. In red we plot the posterior distribution obtained from the full sample of $\{W, \MUV\}$ measurements as described in Section~\ref{sec:bayes}, and infer a neutral fraction of $\xHI = 0.59_{-0.15}^{+0.11}$ ($16-84\%$). The dashed line shows the median value, and shaded region shows the ($16^\mathrm{th}$ and $84^\mathrm{th}$) percentile bounds. For comparison, in blue we plot the posterior for $\xHI$ obtained if we used only the fraction of galaxies emitting \lya\ with $W > 25$\AA, $f_{\mathrm{Ly}\alpha}$. In this case we infer $\xHI = 0.46\pm0.29$. Using the full distribution of EW provides much more information about the evolving IGM compared to $f_{\mathrm{Ly}\alpha}$ and allows for tighter constraints on the neutral fraction.}
\label{fig:posteriorQ_pentericci}
\end{figure}

For each galaxy in this sample, we compute the likelihoods for obtaining the observed equivalent width or upper limit using Equations~\ref{eqn:like_noise} and \ref{eqn:like_limits} for every value of the neutral fraction in our simulations. We exclude the brightest objects ($\MUV < -22$, 1 object) due to the insufficient sampling of very massive halos in the simulations (see Section~\ref{sec:lyart_IGM}) and the uncertainty in their intrinsic EW evolution \citep{Stark2017}, but note this does not affect the inferred neutral fraction for our sample because the UV bright objects are so rare. We use an MCMC sampler \citep{Foreman-Mackey2013} to infer the posterior distribution of $\xHI$ from these data, which is shown in Figure~\ref{fig:posteriorQ_pentericci}. We infer a neutral fraction of $\xHI = 0.59_{-0.15}^{+0.11}$ ($16-84\%$).

\begin{figure}[t] 
\includegraphics[width=0.49\textwidth]{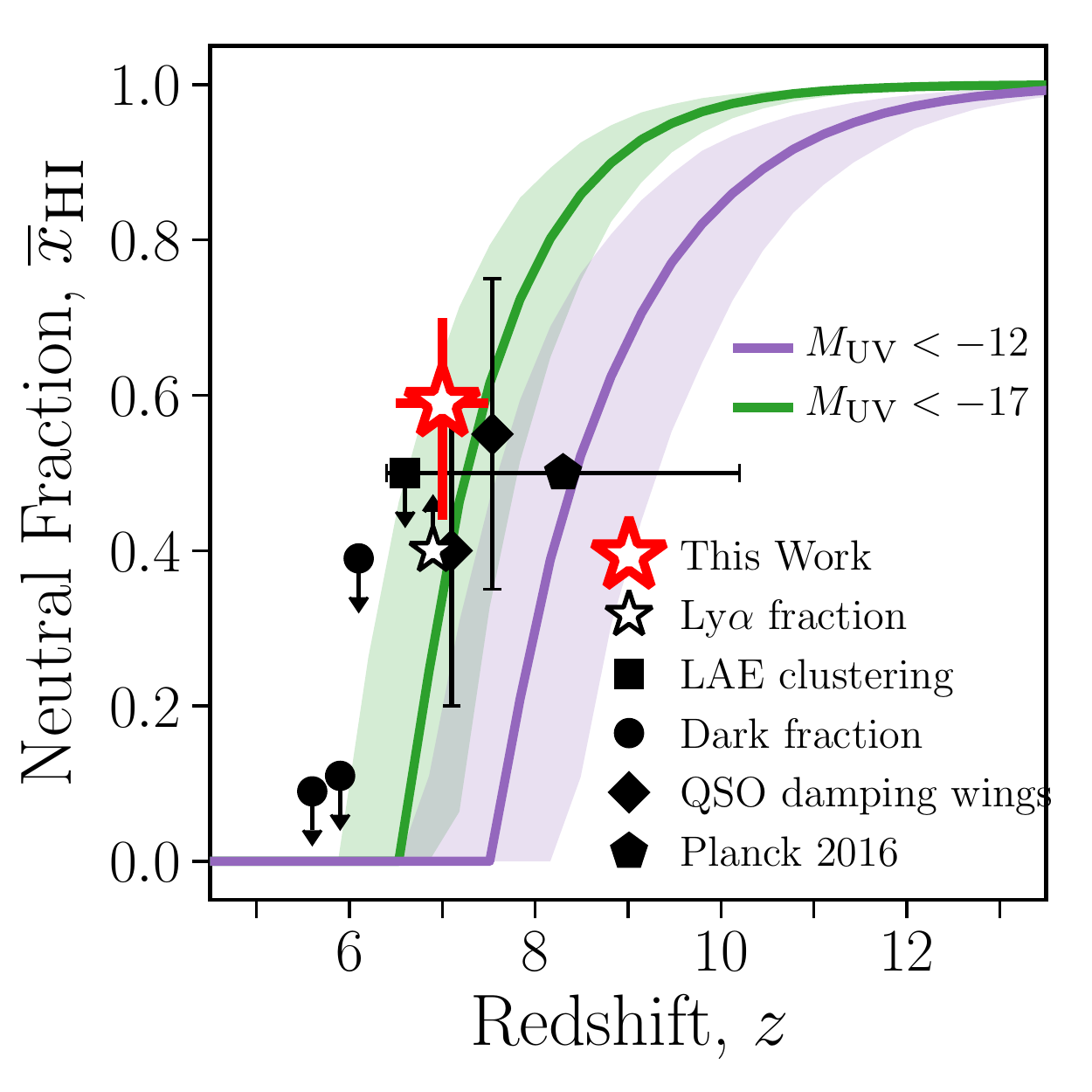}
\caption{The fraction of neutral hydrogen as a function of redshift. Our new constraint is plotted as a red open star. We plot constraints derived from observations of: previous estimates from the fraction of LBGs emitting \lya~\citep[open black star,][]{Mesinger2014}; the clustering of \lya\ emitting galaxies~\citep[square,][]{Ouchi2010,Sobacchi2015a}; \lya\ and Ly$\beta$ forest dark fraction~\citep[circle,][]{McGreer2014}; and QSO damping wings~\citep[diamond,][]{Greig2016,Banados2017}. We offset the constraints at $z\sim7$ \citep[QSO ULASJ1120+0641 damping wing,][\lya\ fraction and our new constraint]{Greig2016b} by $\delta z=0.1$ for clarity. We also plot the \citet{PlanckCollaboration2016a} redshift range of instantaneous reionization (black hatched region). We show as shaded regions the reionization history from integrating the \citet{Mason2015a} UV luminosity function to two magnitude limits of $M_\textsc{uv} = -17$ (green) and $M_\textsc{uv} = -12$ (purple) and drawing from uniform distributions for the ionizing photon escape fraction $10-30$\% and clumping factor $C=1-6$, and log-normal distribution for the ionizing efficiency $\xi_\textrm{ion}$ with mean $25.2$ and standard deviation $0.15$ dex. There are many uncertainties in obtaining the reionization history from luminosity functions so these should not be taken as real constraints on the neutral fraction, but given that galaxies fainter than $M_\textsc{uv} = -17$ likely exist \citep[e.g.,][]{Kistler2009,Weisz2017,Livermore2017,Bouwens2017} our result suggests high escape fractions may not be necessary for reionization.}
\label{fig:Qz}
\end{figure}

This constraint is much tighter than previous measurements of the neutral fraction from \lya\ observations \citep[e.g.,][]{Pentericci2014,Mesinger2014} because we use the full distribution of equivalent widths, $p(W | \MUV)$ in our inference. Previous analyses used only the fraction of galaxies emitting \lya\ with $W > 25$\AA, $f_{\mathrm{Ly}\alpha}$, to constrain the neutral fraction. In Figure~\ref{fig:posteriorQ_pentericci} we also plot the posterior distribution obtained if we had used only $f_{\mathrm{Ly}\alpha}$, i.e. the posterior is $p(\xHI | f_{\mathrm{Ly}\alpha})$, where we compare the simulation $f_{\mathrm{Ly}\alpha}(\xHI)$ derived from Equation~\ref{eqn:like_KDE} with the fraction obtained in \citet{Pentericci2014}: $f_{\mathrm{Ly}\alpha} = 0.29_{-0.15}^{+0.20}$ (for their faint sample, $−20.25 < \MUV < −18.75$). With just the \lya\ fraction we infer a neutral fraction of $\xHI = 0.46\pm0.29$. Clearly, using the full distribution of EW enables us to constrain the neutral fraction much more accurately and, now large samples of LBGs with spectroscopic follow-up are available, should become the statistic of choice for \lya\ reionization inferences.

Where does this constraint sit in our consensus picture of reionization? In Figure~\ref{fig:Qz} we plot constraints derived from observations of: \lya\ emission from galaxies~\citep{Mesinger2014}; the clustering of \lya\ emitting galaxies~\citep{Ouchi2010,Sobacchi2015a}; \lya\ and Ly$\beta$ forest dark fraction~\citep{McGreer2014}; QSO ULASJ1120+0641 damping wings~\citep{Greig2016}. We also plot the neutral hydrogen fraction as a function of redshift, using the \citet{Mason2015a} UV luminosity function model assuming galaxies are the source of ionizing photons and using two limiting magnitudes for the galaxy population: $\MUV < -17$ (currently detectable galaxies) and $\MUV < -12$ (ultra-faint undetected galaxies). The uncertainties in the \citet{Mason2015a} reionization histories comes from the range of possible reionization parameters (e.g., ionizing photon escape fraction, IGM clumping factor, number of ionizing photons per UV photon).

Our constraint is consistent within 1$\sigma$ with the other constraints at $z\sim7$, providing more strong evidence that reionization is on-going at $z\sim7$. Our constraint lies $\Delta \xHI \sim0.2$ higher than the constraint from the $z=7.1$ QSO ULASJ1120+0641 damping wings \citep{Greig2016b}, but is still consistent within the uncertainties.

\subsection{Predictions for JWST}
\label{sec:results_jwst}

\begin{figure}[t] 
\includegraphics[width=0.49\textwidth]{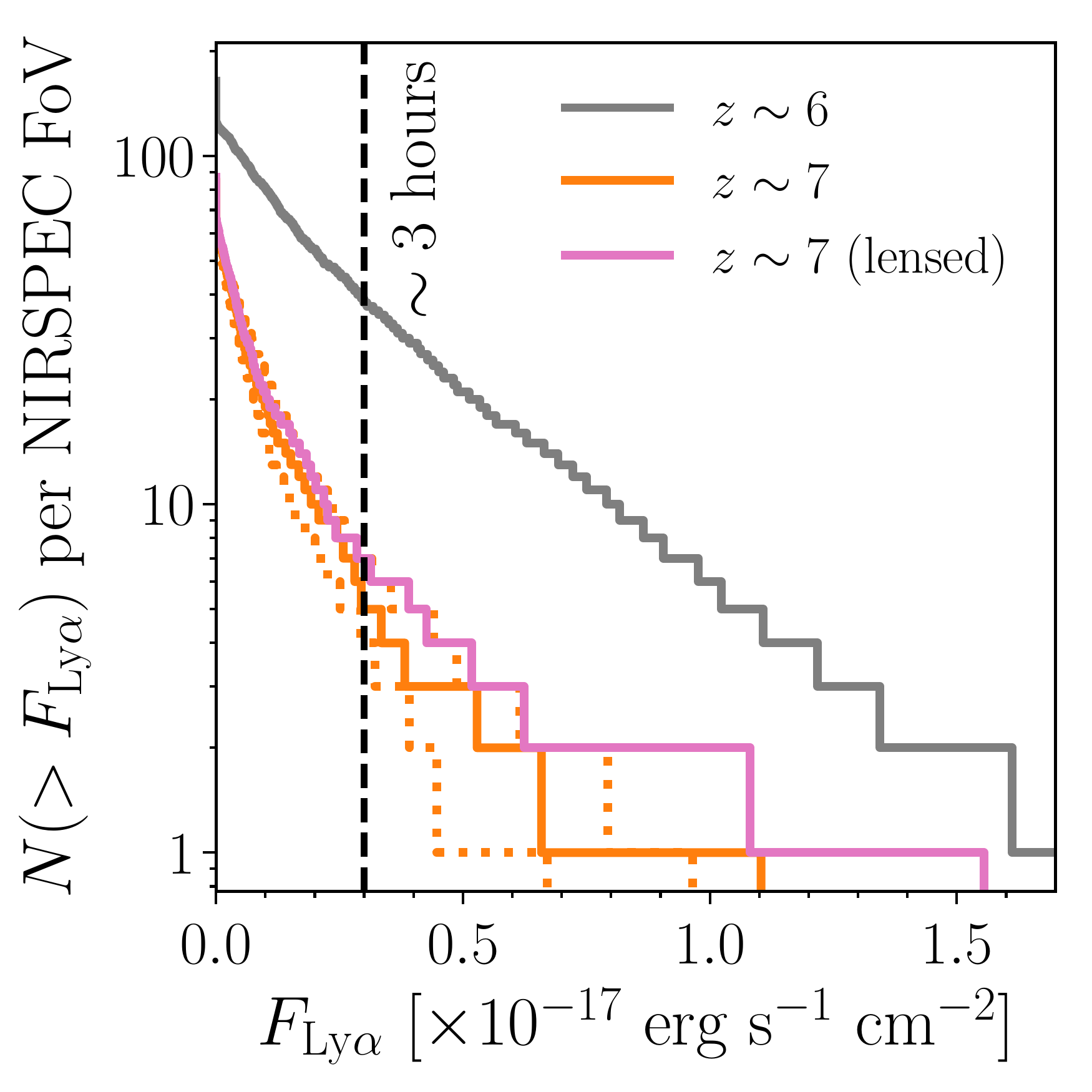}
\caption{Predicted cumulative number counts of LAEs with JWST NIRSpec at $z\sim6$ (gray), and $z\sim7$ (orange) using our recovered neutral fraction $\xHI=0.59_{-0.15}^{+0.11}$. Galaxies are drawn from the \citet{Mason2015a} UV luminosity function model and populated with equivalent widths via $p(W \,|\, \MUV, \xHI)$ - the likelihood described in Section~\ref{sec:bayes_like}. The number counts obtained within the ($16-84\%$) regions on $\xHI$ are shown as dotted orange lines. We also show the cumulative number counts for a gravitationally lensed field where we assume a uniform magnification factor of $\mu=2$ (pink line), which would reveal more emission lines. We obtain the \lya\ fluxes using Equation~\ref{eqn:W_to_flux}. The dashed black line shows the flux limit for a $\sim3$ hour exposure at $R=1000$ with JWST NIRSpec F070LP/G140M at $1-1.5\mu$m calculated with the JWST ETC (\url{https://jwst.etc.stsci.edu})}
\label{fig:JWST_numcounts}
\end{figure}

\begin{figure}[t] 
\includegraphics[width=0.49\textwidth]{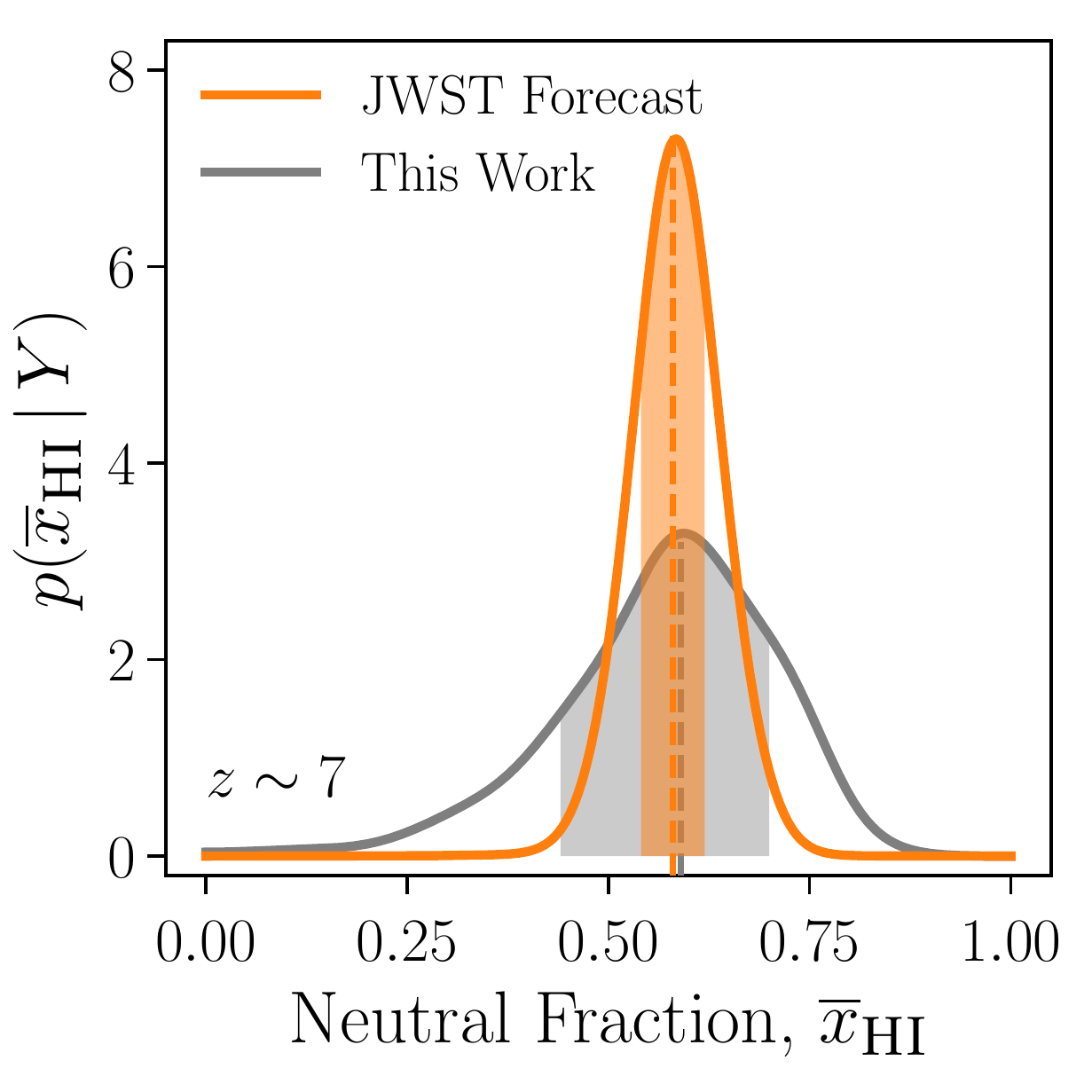}
\caption{Posterior distribution of $\xHI$ for a simulated 10 pointing JWST NIRSpec survey (orange) which is able to tightly constrain the IGM neutral fraction compared to our inference on current observations (red - same as Figure~\ref{fig:posteriorQ_pentericci}). Dashed lines show the median of the distributions, the shaded regions show the $16-84\%$ regions. We take a 10 pointing JWST/NIRSpec mock survey at $z\sim7$ which assumes $\xHI = 0.58$ as described in Section~\ref{sec:results_jwst}, and perform Bayesian inference, assuming a $5\sigma$ flux limit of $> 3\times10^{-18}$ erg s$^{-1}$ cm$^{-2}$. We show the posterior distribution for $\xHI$ inferred from current data (as described in Section~\ref{sec:results_current}) for comparison. In this example JWST could reduce the uncertainty on the neutral fraction by $\sim70\%$.}
\label{fig:JWST_QHI}
\end{figure}

JWST will be uniquely equipped to observe \lya\ and rest-frame optical emission lines into Cosmic Dawn, with extremely sensitive spectrometers NIRSpec and NIRISS covering $1-5 \, \mu$m in a large field of view \citep{Gardner2006,Stiavelli2007}. This will enable direct measurement of the \lya\ $\DV$ and detailed studies of the ISM properties of galaxies during Reionization.

Using our inferred value of $\xHI = 0.59_{-0.15}^{+0.11}$ for the neutral fraction at $z\sim7$ we predict the number of \lya\ emitters detectable in one NIRSpec pointing ($\sim9$ sq arcmins) by drawing galaxies from the \citet{Mason2015a} UV luminosity function and populating them with EW given by our simulated $p(W \,|\, \xHI, \MUV)$. We transform \lya\ equivalent width $W$ to flux using the relation:
\BE \label{eqn:W_to_flux}
f(W, m, z) = W f_0 10^{-0.4m_\textsc{uv}} \frac{c}{\lambda_\alpha^2(1+z)} \left(\frac{\lambda_\textsc{uv}}{\lambda_\alpha}\right)^{-\beta-2}
\EE
where $f_0 = 3.631 \times 10^{-20}$ erg s$^{-1}$ Hz$^{-1}$ cm$^{-2}$, $m_\textsc{uv}$ is the apparent magnitude of the UV continuum, $c$ is the speed of light, $\lambda_\alpha$ is the rest-frame wavelength of \lya, $\lambda_\textsc{uv}$ is the rest-frame wavelength of the UV continuum (usually 1500\AA), and $\beta$ is the UV slope. For simplicity we assume $\beta = -2$, consistent with observations of $z\sim7$ objects \citep[e.g.,][]{Bouwens2012}, though very UV faint galaxies likely have steeper slopes due to extremely low metallicities \citep{Vanzella2016a}.

We plot the predicted number counts in Figure~\ref{fig:JWST_numcounts}, where we assume a 5$\sigma$ UV continuum flux limit of $m_\textsc{ab} > 29$ ($\MUV \sim -18$, corresponding to $\sim1$ hour integration in JWST NIRCam). We predict a 3 hour exposure in one pointing ($\sim 9$ sq arcmins) with JWST NIRSpec will detect $\sim6\pm3$ $z\sim7$ \lya\ lines with a $5\sigma$ flux limit of $\sim 3\times10^{-18}$ erg s$^{-1}$ cm$^{-2}$ (calculated using the JWST ETC), from a total of $\sim80$ LBG dropout detections. We also show the forecast for a cluster lensing survey \citep[e.g., GLASS,][]{Treu2015,Schmidt2016} assuming a simple uniform magnification factor of $\mu=2$ due to gravitational lensing (i.e. $p(\mu) = \delta(\mu - 2)$). In this case, all fluxes are magnified by $\mu$ whilst the area decreases by $1/\mu$, and assuming the same flux limit as above we predict $\sim10\pm2$ \lya\ lines from a total of $\sim90$ LBG detections. The NIRSpec field-of-view is still small compared to large scale structure at $z\sim7$ so wide area random pointing surveys will be essential to estimate the global $\xHI$. 

We simulate a 10 pointing NIRSpec survey with F070LP/G140M ($R=1000$), with 3 hour exposures in each field, by again sampling the \citet{Mason2015a} luminosity function in a larger area. We perform the inference on these mock JWST observations at $z\sim7$, assuming $\xHI=0.59$. This yields $\sim60$ detections from $\sim800$ LBGs. Again, we assume a $5\sigma$ flux limit of $> 3\times10^{-18}$ erg s$^{-1}$ cm$^{-2}$. The posterior distribution obtained from the JWST mock observations is shown in Figure~\ref{fig:JWST_QHI}, with the posterior from the current observations (Section~\ref{sec:results_current}) shown for comparison. We obtain $\xHI=0.60_{-0.06}^{+0.02}$, a $\sim70\%$ reduction in uncertainty compared to the current sample. We note this is an average forecast, and a more realistic survey forecast would require sampling the simulation directly \citep[e.g.,][]{Mesinger2008}. We also caution our mock survey assumes 100\% completeness, and maximized filling of NIRSpec slits, but, nevertheless, observations with NIRSpec will constrain the neutral fraction much more tightly than current observations. 

\section{Discussion}
\label{sec:dis}

In this section we discuss our result in the context of other probes of reionization (Section~\ref{sec:dis_history}), and we discuss the implications of the mass-dependent \lya\ velocity offset on the evolving \lya\ fraction for average (Section~\ref{sec:dis_DV}) and UV bright (Section~\ref{sec:dis_UVbright}) galaxies.

\subsection{The global reionization history}
\label{sec:dis_history}

Robust constraints on the reionization history are challenging. Whilst quasars provide high S/N information about individual (but rare) lines of sight they are likely to be biased to overdense and more ionized regions \citep{Barkana2004,Mesinger2010,Decarli2017}, and the number densities of bright quasars drop dramatically at $z>6$ \citep{Fan2001,Manti2016,Parsa2017}. Constraining reionization with large samples of galaxies clearly avoids these problems; with the help of gravitational lensing in clusters, e.g. the Frontier Fields \citep{Lotz2016}, we know there are large populations of faint galaxies at $z>6$ \citep{Yue2014,Atek2015,Livermore2017,Vanzella2017}, and GRB host galaxy searches indicate far fainter galaxies must also exist \citep{Kistler2009,Trenti2012a}.

\lya\ emission from galaxies has long been recognized as a probe of reionization \citep{Haiman1999,Malhotra2004,Santos2004,Verhamme2006,McQuinn2007,Dijkstra2014a}, and the framework presented in this paper provide a direct constraint on the IGM neutral fraction from observations of \lya\ emission from galaxies, incorporating both realistic galaxy properties and realistic IGM topologies for the first time.

Our constraint on the neutral fraction, $\xHI = 0.59_{-0.15}^{+0.11}$, is consistent with other robust probes of IGM neutrality at $z\sim7$ \citep{Mesinger2014,Greig2016b} demonstrating the power of \lya\ follow-up of LBGs to constrain the neutral fraction, and providing more strong evidence that the IGM is undergoing significant reionization at $z\sim7$. Using the full distribution of observed $\{W, \MUV\}$ as inputs to our inference provides much tighter constraints than using the standard `\lya\ fraction', as we demonstrated in Figure~\ref{fig:posteriorQ_pentericci}.

Our median value lies $\Delta \xHI \sim0.2$ higher than that inferred by \citet{Greig2016b} from the QSO ULASJ1120+0641 damping wings at $z=7.1$, which was obtained using the same IGM simulations, though our posterior distribution is marginally skewed to lower values (see Figure~\ref{fig:posteriorQ_pentericci}). This offset is not significant given the uncertainties, and does not require us to invoke any additional evolution in galaxy properties. Within the next few years larger samples, as demonstrated in our mock survey with JWST described in Section~\ref{sec:results_jwst}, will greatly reduce the uncertainties in our constraints from \lya\ detections and non-detections.

With large samples, it will be possible to measure the variations in $\xHI$ over the sky, and cross-correlate with other constraints from quasars and eventually 21cm observations \citep{Lidz2009,Vrbanec2016,Sobacchi2016,Mirocha2016,Mesinger2016,Greig2017} to directly observe the inhomogeneous process of reionization. Furthermore, with tighter constraints on the timeline of reionization, it will be possible to better constrain the sources of ionizing photons: as the ionizing photon budget from galaxies depends on e.g., the minimum mass/luminosity of galaxies and the rate of ionizing photons per unit UV luminosity.

\subsection{A sudden drop in \lya\ emission \\ -- redshift evolution of $\DV$?}
\label{sec:dis_DV}

In our model, we include empirically calibrated relations for both the intrinsic dependence of \lya\ EW on UV magnitude and ISM radiative transfer in galaxies of a given halo mass (UV magnitude at fixed redshift), which builds in a simple redshift evolution assuming galaxies of the same UV magnitude live in less massive halos at higher redshifts. In this framework, UV faint galaxies have intrinsically high EW than UV bright galaxies and lower \lya\ velocity offsets.

These correlations are motivated by numerous observations of \lya\ emission from galaxies at a range of redshifts, including very low redshift samples where detailed spatial and spectral observations are possible \citep{Hayes2013,Yang2016}. It is likely the density and distribution of neutral gas in the ISM plays a key role in the mediation of \lya\ propagation through galaxies: an ISM with high column densities of neutral hydrogen, $N_\textsc{hi}$, scatters \lya\ photons more significantly, spectrally and spatially \citep{Verhamme2006,Zheng2010}. Observations of $z<4$ galaxies confirm high $N_\textsc{hi}$ correlates with high \lya\ velocity offset \citep{Yang2016,Hashimoto2015,Henry2015}, and more \lya\ extended halos \citep{Hayes2013,Guaita2017}. 

With increasing redshift, when galaxies were less massive \citep{Lacey1993}, \lya\ should escape more easily with high EW: these galaxies will contain less dust \citep[as ALMA and Plateau de Bure Interferometer (PdBI) results are suggesting,][]{Walter2012,Ouchi2013,Ota2014,Schaerer2015,Maiolino2015,Capak2015,Bouwens2016,Pentericci2016} and neutral gas than at low redshifts. Additionally, the covering fraction of neutral hydrogen may evolve with galaxy mass, star formation rate, stellar populations and/or redshift. Hard ionizing spectra from low metallicity stars \citep[which may be significant at high redshifts,][]{Mainali2016,Stark2015,Schmidt2017,Stark2017} can create more ionized holes through the ISM, reducing the covering fraction, an effect which is enhanced for low mass galaxies \citep{Trebitsch2017}. A low covering fraction would facilitate \lya\ escape closer to the galaxy systemic velocity, and some observations have indicated a decreasing covering fraction with redshift \citep{Leethochawalit2016,Jones2013}.

All these factors, and the correlation of $\DV$ with halo mass as shown in Figure~\ref{fig:DV_Muv_lit}, suggest velocity offsets should decrease with increasing redshift for galaxies at fixed UV magnitude. These low velocity offsets are correlated with reduced scattering within the ISM and thus a higher EW of \lya. This should increase the visibility of \lya\ \textit{until} the IGM starts to become neutral and these low $\DV$ lines are easily attenuated by nearby neutral hydrogen. As was noted by \citet{Mesinger2014} and \citet{Choudhury2015} this offers a simple explanation for evolving galaxy properties which may accelerate the decline in \lya\ in UV faint galaxies.

\subsection{\lya\ from UV bright galaxies \\-- redshifted away from resonance?}
\label{sec:dis_UVbright}

A high fraction of \lya\ observed in some UV bright ($\MUV < -21.5$) galaxies at $z > 6$ \citep[][though c.f. \citet{Treu2013} for non-detections of \lya\ in slightly fainter galaxies]{Curtis-Lake2012,Stark2017} is surprising for several reasons: the electron scattering optical depth from the \citet{PlanckCollaboration2016a} favors a significant IGM neutral fraction at these redshifts, with instantaneous reionization occurring at $z=7.8 - 8.8$; and the observed fraction of UV faint galaxies appears to steadily decrease at the same redshifts \citep{Pentericci2014,Schenker2014}. Why can we more easily see \lya\ in some UV bright galaxies into the Epoch of Reionization?

The most UV bright galaxies at high redshift probably reside in halos with mass $\simgt10^{11} \msun$, which may already have stable gaseous disks, as suggested by recent ALMA observations of two UV bright galaxies at $z\sim6$ \citep{Smit2017a} and observations of stable rotation in low mass galaxies at $z\sim1-2$ \citep{Stott2016,Mason2016}. Thus, it is likely \lya\ photons traveling from these galaxies will experience significant radiative transfer effects with the ISM. 

The enhanced Doppler shift of the emerging \lya\ photons in UV bright galaxies provides some explanation for the high fraction of \lya\ observations for these populations compared to UV faint galaxies at $z\sim7$. As shown in Figures~\ref{fig:T_Muv} and \ref{fig:like}, we predict transmission of UV bright galaxies evolves more slowly with the evolving IGM compared to fainter objects, making them visible far into the epoch of reionization and thus prime targets for spectroscopic confirmation. Though note their underlying EW distribution is likely much steeper and has a higher peak of non-emitters than for UV faint galaxies. When \lya\ is emitted from UV bright objects it is likely to have low EW as the photons are so dispersed spatially and spectrally. 

However, this effect is also highly correlated to the large scale environment in which these galaxies reside; assessing the relative contributions of evolving galaxy properties and environment to this apparent increase in the \lya\ fraction is explored by \citet{Mason2018}. The high \lya\ transmission of UV bright galaxies make them ideal targets for spectroscopic follow-up to understand the star formation processes occurring in the early universe.

\section{Summary and Conclusions}
\label{sec:conc}

We have developed a flexible Bayesian inference framework to infer the IGM neutral fraction during reionization by forward-modeling the observed equivalent width distribution of \lya\ emission from LBGs. Our model incorporates sightlines through realistic IGM simulations to model galaxies with realistic ISM properties.

Our main conclusions are as follows:

\begin{enumerate}[(i)]
\item The \lya\ line profile emerging from the ISM has a huge impact on the probability of transmission through the IGM \citep{Dijkstra2011}, and is related to the properties of the source galaxy. This must be systematically accounted for in reionization inference.
\item We introduce a simple empirical relation between the halo mass of a galaxy (or UV luminosity at fixed redshift) and its \lya\ line peak velocity offset, where the most massive galaxies have the largest velocity offsets likely due to increased $N_\textsc{HI}$ in the ISM, higher halo circular velocities and/or the presence of star-formation induced outflows.
\item This relation predicts that with increasing redshift, \lya\ velocity offsets will decrease for galaxies at fixed UV luminosity, making \lya\ lines more susceptible to absorption in the IGM. This effect would accelerate the decline in \lya\ emission compared to other reionization probes and be a factor in explaining the sudden drop of \lya\ emission observed at $z>6$.
\item We conduct a Bayesian inference from current observations at $z\sim7$ from \citet{Pentericci2014} and infer the first direct constraint on the neutral fraction from \lya\ transmission of $\xHI = 0.59_{-0.15}^{+0.11}$, which is consistent with other robust probes of the neutral fraction and confirms reionization is on-going at $z\sim7$.
\item Using the full distribution of \lya\ equivalent width measurements enables us to provide much tighter constraints on the neutral fraction compared to the standard `\lya\ fraction', $P(W > 25\textrm{\AA})$, used in previous analyses.
\item We make predictions for spectroscopic surveys with JWST and find a $\sim30$ hour LBG follow-up survey with JWST/NIRSpec could reduce the uncertainty in $\xHI$ by $\sim70\%$.
\end{enumerate}

Future near-IR spectrographs in space, such as JWST NIRSpec and NIRISS, will be able to observe both \lya\ and rest-frame optical lines for galaxies to $z\simlt12$ and to measure SFRs and \lya\ velocity offsets for these objects, enabling us to further understand the interactions between star-forming regions, the ISM, and the reionizing IGM. It will soon be possible to apply our framework to large samples, free of cosmic variance, to get accurate universal constraints on the evolution of the neutral fraction.

\acknowledgments

The authors thank Dawn Erb and Dan Stark for providing their observational data. We thank Simon Birrer, Fred Davies, Max Gronke, Joe Hennawi and Crystal Martin for useful discussions. 

C.M. acknowledges support by NASA Headquarters through the NASA Earth and Space Science Fellowship Program Grant NNX16AO85H. A.M. acknowledges support from the European Research Council (ERC) under the European Unions Horizon 2020 research and innovation program (grant agreement No 638809 AIDA). This research was partially supported by the Australian Research
Council through awards FT130101593 and CE170100013. This work was supported by the HST BoRG grants GO-12572, 12905, and 13767, and the HST GLASS grant GO-13459

This work made use of the following open source software: IPython \citep{Perez2007}, matplotlib \citep{Hunter2007}, NumPy \citep{VanderWalt2011}, SciPy \citep{Oliphant2007}, Astropy \citep{AstropyCollaboration2013} and EMCEE \citep{Foreman-Mackey2013}.

\bibliography{library,apj-jour}
\bibliographystyle{yahapj}

\end{document}